\documentclass[]{aa}
\makeatletter
\renewcommand*\aa@pageof{, page \thepage{} of \pageref*{LastPage}}
\makeatother
\newcommand{\cha}{\textit{Chandra}}

\def\flu{{erg\,s$^{-1}$\,cm$^{-2}$}}

\def\aap{A\&A}

\def\apjl{ApJL}
\def\apjs{ApJS}

\usepackage{natbib}
\usepackage{float}
\usepackage{color}
\usepackage{graphicx}
\usepackage{textcomp,gensymb}
\usepackage{array,amsmath}
\usepackage{mathtools}
\usepackage{multirow,makecell}
\usepackage{enumitem}
\usepackage{longtable}
\usepackage{hyperref}
\begin{document}

\title{LBT-MODS spectroscopy of high-redshift candidates in the \cha\ J1030 field. A newly discovered $z\sim$2.8 large scale structure
}
\titlerunning{LBT-MODS follow-up of High-z X-ray AGN in the J1030 field}

\author{Stefano~Marchesi\inst{1,2,3} \and Marco~Mignoli\inst{1} \and Roberto~Gilli\inst{1} \and Giovanni~Mazzolari\inst{3,1} \and Matilde~Signorini\inst{4,5} \and Marisa~Brienza\inst{1,3} \and Susanna~Bisogni\inst{6} \and Micol~Bolzonella\inst{1} \and Olga~Cucciati\inst{1} \and Quirino~D'Amato\inst{7} \and Alessandro~Peca\inst{8} \and Isabella~Prandoni\inst{9} \and Paolo~Tozzi\inst{5} \and Cristian~Vignali\inst{3,1} \and Fabio~Vito\inst{1} \and Andrea~Comastri\inst{1}
}

\institute{INAF - Osservatorio di Astrofisica e Scienza dello Spazio di Bologna, Via Piero Gobetti, 93/3, 40129, Bologna, Italy 
\and Department of Physics and Astronomy, Clemson University, Kinard Lab of Physics, Clemson, SC 29634, USA 
\and Dipartimento di Fisica e Astronomia, Università degli Studi di Bologna, via Gobetti 93/2, 40129 Bologna, Italy 
\and Dipartimento di Fisica e Astronomia, Università degli Studi di Firenze, via G. Sansone 1, 50019 Sesto Fiorentino, Firenze, Italy 
\and Istituto Nazionale di Astrofisica (INAF) -- Osservatorio Astrofisico di Arcetri, Largo E. Fermi, I-50125 Firenze, Italy
\and INAF -- Istituto di Astrofisica Spaziale e Fisica Cosmica Milano, Via A. Corti 12, I-20133 Milano, Italy
\and Scuola Internazionale Superiore di Studi Avanzati (SISSA), Via Bonomea 265, 34136, Trieste, Italy
\and Department of Physics, University of Miami, Coral Gables, FL 33124, USA
\and INAF -- Istituto di Radioastronomia, via P. Gobetti 101, 40129 Bologna, Italy 
}

\abstract{We present the results of a spectroscopic campaign with the Multi-Object Double Spectrograph (MODS) instrument mounted on the Large Binocular Telescope (LBT), aimed at obtaining a spectroscopic redshift for seven \cha\ J1030 sources with a photometric redshift $\geq$2.7 and optical magnitude $r_{AB}$=[24.5--26.5]. We obtained a spectroscopic redshift for five out of seven targets: all of them have $z_{\rm spec}\geq$2.5, thus probing the reliability of the \cha\ J1030 photometric redshifts. The spectroscopic campaign led to the serendipitous discovery of a $z\sim$2.78 large scale structure (LSS) in the J1030 field: the structure contains four X-ray sources (three of which were targeted in the LBT-MODS campaign) and two non-X-ray detected galaxies for which a VLT-MUSE spectrum was already available. We also find 26 galaxies with a photometric redshift in the range $z_{\rm phot}$=[2.68--2.88], which we consider candidate LSS members. 
The X-ray members of the LSS are hosted in galaxies that are significantly more massive (log(M$_*$/M$_{\odot}$)=[10.0--11.1]) than both those hosting the two MUSE-detected sources (log(M$_*$/M$_{\odot}$)$<$10) and those belonging to the photometric sample (median stellar mass log(M$_*$/M$_{\odot}$)=10.0). Both observations and simulations show that massive galaxies, and particularly objects having log(M$_*$/M$_{\odot}$)$>$10, are among the best tracers of large scale structures and filaments in the cosmic web. Consequently, our result can explain why X-ray-detected AGN have also been shown to be efficient tracers of large scale structures.
}


\maketitle

\section{Introduction}

The formation and evolution of galaxy clusters is a widely debated topic, involving multiple complex processes. These include, for example, merging and interaction of gas-rich galaxies, growth of supermassive black holes at galaxy centers, and the consequent AGN feedback (either positive or negative) on the intra-cluster medium and on the star formation (SF) processes \citep[e.g,][]{overzier16}. In gas-rich, overdense regions, evolutionary processes likely take place faster and/or more efficiently than in field galaxies, as shown through measurements of the mass-metallicity relation at $z\sim$0 \citep[e.g.,][]{peng14}. Furthermore, several works have shown that at $z>$2 the X-ray AGN activity is enhanced in sources in protoclusters with respect to those in the field \citep[e.g.,][]{pentericci02,digby10,lehmer13,martini13,vito20, tozzi22}, and a similar trend is also observed in terms of star-formation activity \citep[e.g.,][]{capak11,umehata15,umehata19,gilli19,kubo19,chapman23}. The study of large scale structures at $z>$2 hence represents an ideal way to investigate all of these processes, and understand the evolving path that leads to $z\sim$0 galaxy clusters.

Cosmological simulations show that only a minor part of the whole structure mass lies within the inner part of a protocluster, in the so-called main halo, which is the most massive among the many progenitor halos that will eventually merge in the $z\sim$0 cluster \citep[see, e.g.,][]{muldrew15}. Indeed, at $z\sim$2 about 80\,\% of the protocluster mass and member galaxies are expected $-$ on average $-$ to reside outside the main halo, over a scale of several Mpc. \citet{muldrew15} also showed that this behavior is mass--dependent, meaning that the progenitors of more massive clusters extend over larger scales. For this reason, a full understanding of the physics and evolution of a large-scale structure requires tracking sources over an area of several Mpc$^2$ \citep[i.e., at $z\sim$2--3, over field of views of at least $\sim$15$^\prime\times$15$^\prime$; see also][]{chiang13}. To this end, faint X-ray selected AGN have been shown to be excellent tracers of large scale structures \citep[LSSs; see, e.g.,][]{gilli03,silverman10,marchesi21}. For example, \citet{gilli03} identified several spikes in the redshift distribution of X-ray sources in the \textit{Chandra} Deep Field-South, some of which containing 3-4 X-ray sources only. All of these LSSs were eventually confirmed with extended photometric and spectroscopic campaigns \citep{castellano07,salimbeni09,balestra10}.

 Recently, \cha\ targeted with a deep, $\sim$0.5\,Ms observation a 335\,arcmin$^2$ region centered on the $z$=6.31 quasar SDSS J1030+0525.  SDSS J1030+0525 is one of the first quasars detected at {\it z}$\,>\,$6 by the Sloan Digital Sky Survey \citep[SDSS;][]{fan01}. This field is known to be highly biased, since it hosts both a galaxy overdensity at $z$=6.3, around the SDSS J1030+0525 \citep{morselli14,balmaverde17,mignoli20}, and another overdensity at $z$=1.7, around a Fanaroff-Riley type II (FRII) radio galaxy \citep{nanni18,gilli19,damato20}. The flux limit achieved by the \cha\ J1030 survey \citep{nanni20} makes this the fifth deepest X-ray field to date: the survey contains 256 sources detected down to a 0.5--2\,keV flux limit f$_{\rm 0.5-2\,keV}$=6$\times$10$^{-17}$\,\flu. Such a deep flux limit makes J1030 an ideal region to search for intrinsically faint and/or heavily obscured AGN. Optical/near-infrared (nIR) counterparts selected in the $r$, $z$, J and 4.5\,$\mu$m were associated with the X-ray sources using standard likelihood-ratio matching techniques. Spectroscopic and photometric redshifts have been computed in \citet[][hereafter M21]{marchesi21} for 243 extragalactic sources out of the 256 \cha\ J1030 ones\footnote{Out of the remaining 13 objects, seven are stars and six lack an optical/nIR counterpart and cannot therefore be classified.}. As reported in M21, X-ray sources appear to efficiently trace large scale structures in the \cha\ J1030 field too. In fact, multiple significant X-ray structures are detected in the field in the redshift range $z$=[0.15--1.5], all of which were confirmed (mostly with increased significance) when extending the analysis to non-X-ray sources.
 
 The spectroscopic completeness of deep X-ray surveys, while fairly high when considering samples in their entirety (up to $\sim$50--70\,\%; see, e.g., \citealt{luo17} on the \cha\ Deep Field-South 7\,Ms survey; \citealt{marchesi16a} on the \cha\ COSMOS-Legacy survey; M21 on the \cha\ J1030 survey), is significantly biased against high-redshift sources. In particular, at redshift $z>$3 (i.e., at epochs before the peak in AGN and star formation activity, see, e.g., \citealt{madau14,ueda14,buchner15,ananna19,peca23}) only a few hundred X-ray selected AGN from X-ray surveys have been spectroscopically confirmed\footnote{This number does not include those objects, mostly unobscured Type 1 quasars, that were detected at other wavelengths and then followed up with X-ray facilities \citep[see, e.g.,][]{nanni17,vito19}}. Such a result is not surprising, given that more distant sources are expected to be fainter and therefore more difficult to be identified spectroscopically.  In the \cha\ J1030 survey, while a spectroscopic redshift was measured for 123 out of 256 sources, only three of them have $z_{\rm spec}>$3, one of which being the $z$=6.3 QSO at the center of the J1030 field.

Within this framework, we present here the results of a spectroscopic campaign with the Multi-Object Double Spectrograph (MODS) mounted on the Large Binocular Telescope (LBT) which targeted \cha\ J1030 sources with a photometric redshift $z_{\rm phot}\geq$2.7. The paper is organized as follows: in Section~\ref{sec:new_data} we report on new spectroscopic and photometric observations of the J1030 field we recently performed. In Section~\ref{sec:spectra_results} we then present the results of the LBT-MODS spectroscopic campaign, while in Section~\ref{sec:z_28_LSS} we use the new spectroscopic redshifts to report on the discovery of a $z\sim$2.8 large scale structure in the J1030 field. We finally summarize the results of the paper in Section~\ref{sec:summary}. In Appendix~\ref{sec:logn-logs} we also present the updated photometric redshifts for the $z>$2 \cha\ J1030 sources which we computed using new LBT Large Binocular Cameras (LBC) $g$ band and Wide-field InfraRed Camera (WIRCAM)-CFHT Ks band photometry, and use the new redshifts to update the \cha\ J1030 $z>$3 number counts

Through the rest of the work, we assume a flat $\Lambda$CDM cosmology with H$_0$ = 69.6\,km\,s$^{-1}$\,Mpc$^{-1}$, $\Omega_m$=0.29 and $\Omega_\Lambda$=0.71 \citep{bennett14}. Errors are quoted at the 90\,\% confidence level, unless otherwise stated.

\section{New spectroscopic and photometric coverage of the J1030 field}\label{sec:new_data}
Among the 26 \cha\ J1030 sources with photometric redshift $z_{\rm phot}\geq$2.7 and no spectroscopic redshift available in the M21 catalog, we selected seven targets that were bright enough ($r_{AB}\lesssim$26.5) to obtain an optical spectrum with LBT-MODS. Since we used a multi-object spectrograph like MODS, we also included 12 more \cha\ J1030 sources that were randomly located inside the footprint of the two masks and lacked a spectroscopic redshift. These additional targets had photometric redshift in the range $z_{\rm phot}$=[0.9--2.2]. 

Our LBT-MODS observations were performed on January 31 and February 5, 2022: each of the two masks was observed for a total of four hours. Since LBT is made of two telescope units of 8.4m diameter each that are equipped with similar instrumentation (and in particular with the identical MODS1 and MODS2 spectrographs), the total time on target is actually halved.
Our MODS spectra have been obtained in dichroic mode using both the blue G400L (3500$-$5900\,\AA) and the red G670L (5400$-$10000\,\AA) gratings on the blue and red channels. As shown in the following paragraphs, the wide wavelength range we obtained by using this specific combined configuration (the same adopted in M21) made it possible to detect emission and absorption features over a wide range of wavelengths and to reliably measure a significant number of spectroscopic redshifts. We reduced the data using the spectroscopic pipeline to reduce optical/near-infrared data from slit-based spectrographs SIPGI \citep{bisogni22,gargiulo22}.

The data reduction pipeline is organized as follows. We create calibration frames for both MODS1 and MODS2 for each of the two sets of observations (i.e., for each of the two masks). Specifically, a bad pixel map is produced with imaging flats and applied to every observed frame, along with a correction for cosmic rays. Each frame is independently bias subtracted and flat-field corrected through a master flat obtained from a set of spectroscopic flats. The inverse solution of the dispersion, obtained from a set of arc lamps frames and stored in the master lamp, is applied to individual frames for the wavelength calibration and the removal of any curvature due to optical distortion. The mean accuracy we achieved on the wavelength calibration is 0.09\AA\ for the red and 0.07\AA\ for the blue channel. 2D wavelength-calibrated spectra are then extracted and sky subtracted. The sensitivity function, obtained thanks to the observations of the spectro-photometric standard stars GD153 and GD71, is applied to obtain flux-calibrated spectra. Finally, wavelength- and flux-calibrated, sky-subtracted spectra are stacked together and the 1D spectrum of each source is extracted.

The targets redshifts have been measured on the 1D spectra: for the MODS observations, the wavelength coverage allows a Ly$\alpha$ detection down to $z=1.9$, where the Ly$\alpha$ emission line is often the only detectable feature for our faint objects. All the emission line centers were determined using a Gaussian fit to the line profile: the only exception was the Ly$\alpha$ feature of XID~011 where, due to the structured shape, we adopted the line peak to determine the redshift. For all the solutions for which the single emission line was identified as Ly$\alpha$, we can confidently rule out other possible redshift solutions. Lower redshift solutions would imply the detection of other nearby lines, as is the case for the X-ray sources with $z<$1.5 presented in Appendix~\ref{sec:app_spectra}. Also, in all the objects belonging to the $z\sim$2.78 structure the emission line is asymmetric with a red wing and, when the spectra signal-to-noise allows it, the underlying continuum shows a clear break at the wavelength of the  Ly$\alpha$ line. Both spectral characteristics are typical of the high-redshift Ly$\alpha$ line.
 
Our group was also recently granted time to observe the J1030 field with LBT-LBC in the $g$ band (observations performed in May, 2021; total exposure 1800\,s; PI M. Mignoli) and with WIRCAM-CFHT in the Ks band (observations performed in February, 2022; total exposure 6400\,s; PI M. Mignoli). We will present a detailed description of the data analysis process, as well as an updated photometric catalog, in M. Mignoli et al. (in prep.). Here we briefly summarize the improvement provided by these observations with respect to the photometric catalog we used in M21. 

The $g$ band imaging (3\,$\sigma$ magnitude limit $g_{\rm AB}$=27.5) covers a band that was previously missing in our photometric coverage of the field, and its depth almost matches the depth we achieved in the adjacent $r$ band (magnitude limit $r_{\rm AB}$=27.5).
The J1030 field was instead already observed in the Ks band as part of the MUSYC survey. However, our new Ks data is $\sim$3 magnitudes deeper (3\,$\sigma$ magnitude limit Ks$_{\rm AB}$=24) than the MUSYC K-Wide \citep{blanc08} image we used in M21, and $\sim$1 magnitude deeper than the MUSYC K-Deep coverage of the central 10$^\prime\times$10$^\prime$ of the J1030 field \citep{quadri07}. In Section~\ref{sec:z_28_LSS}, we take advantage of this new photometric information to compute the stellar masses and star formation rates of the galaxies belonging to a newly detected $z\sim$2.78 large scale structure in the J1030 field, as well as the photometric redshifts of non X-ray sources in the J1030 field, searching for more candidate members of the same structure. In Appendix~\ref{sec:logn-logs}, we instead show how we used the data from this imaging campaign to refine the \cha\ J1030 photometric redshifts and to derive the $z>$3 X-ray number counts in the J1030 field.

\section{Results of the LBT-MODS spectroscopic campaign}\label{sec:spectra_results}

We report in Table~\ref{tab:redshift_summary} the redshift values we obtained, together with a breakdown of the different features detected in the spectra. We measured a spectroscopic redshift for five out of seven sources in our main sample of high-redshift candidates: all of them have 2.5$<z_{\rm spec}<$3.1. No clear feature was instead detected in the spectra of XID 378 and XID 381. As reported in Table~\ref{tab:redshift_summary}, the photometric redshifts computed in M21\footnote{All high-$z$ candidates but XID 381 have been observed in at least 11 photometric bands. XID 381 has been observed in 9 bands.} efficiently selected high--redshift candidates: all five $z\geq$2.7  candidates for which we obtained a spectroscopic redshift have $z_{\rm spec}\geq$2.5. This increases the \cha\ J1030 spectroscopic completeness at $z>$2.5 (from 8 to 13 out of 40 \cha\ J1030 sources with $z>$2.5, i.e., from 20\,\% to  33\,\%) and shows that the \cha\ J1030 photometric redshifts can reliably select high-redshift candidates. We also measured a spectroscopic redshift for 7 out of the 12 \cha\ J1030 filler sources at 0.9$<z<$2.2 lacking a spectroscopic redshift.

\begingroup
\renewcommand*{\arraystretch}{1.5}
\begin{table*}
\centering
\scalebox{0.82}{
\vspace{.1cm}
 \begin{tabular}{cccccccccc}
 \hline
  \hline
ObjID &    RA    &    DEC   & $r_{\rm AB}$ & $z_{\rm spec}$ & $Q_{\rm z}$ & Class & $z_{\rm phot, M21}$ & $z_{\rm phot, new}$    &  Spectral features (obs $\lambda$) \\ 
      & hh:mm:ss & dd:mm:ss \\
 \hline
XID 007 & 10:30:29.77 & +05:22:32.26 & 25.21 & 2.7799   & 2 & ELG & 3.24$_{-1.30}^{+1.30}$ & 3.22$_{-1.28}^{+1.28}$  &  Ly$\alpha$ (4595\,\AA) \\  
XID 008 & 10:30:28.10 & +05:22:58.75 & 24.41 & 2.7797  & 2 & ELG&  3.34$_{-0.90}^{+0.90}$ & 2.68$_{-1.02}^{+1.02}$ &  Ly$\alpha$ (4597\,\AA), NV \\ 
XID 011 & 10:30:27.89 & +05:23:13.42 & 26.67 & 2.7674  & 2 & ELG & 4.34$_{-2.10}^{+2.10}$ & 4.16$_{-1.48}^{+1.48}$ &  Ly$\alpha$ (4581\,\AA) \\
XID 151 & 10:30:41.54 & +05:26:25.13 & 25.25 & 3.041   & 2 & ELG & 2.74$_{-0.88}^{+0.88}$ & 2.96$_{-0.52}^{+0.52}$ & Ly$\alpha$ (4912\,\AA)  \\ 
XID 158 & 10:30:28.01 & +05:28:08.35 & 24.83 & 2.505  & 1 & ELG & 3.37$_{-1.17}^{+1.19}$ & 2.12$_{-0.88}^{+0.88}$   &  Ly$\alpha$ (4262\,\AA), NV \\
XID 378 & 10:30:38.13 & +05:24:03.31 & 26.5  & --  & 0 & None &  5.45$_{-1.05}^{+1.03}$  & 1.92$_{-0.04}^{+2.82}$ & None \\ 
XID 381 & 10:30:43.12 & +05:25:26.99 & 25.06 & -- &  0 & None   & 3.44$_{-1.56}^{+1.56}$ & 3.86$_{-1.00}^{+1.00}$ & None \\
\hline
\hline
XID 029 & 10:30:38.00 & +05:26:13.09 & 26.66 & --  & 0 & None  & 2.22$_{-0.80}^{+0.80}$  & 2.15$_{-0.77}^{+2.11}$  &   None  \\ 
XID 066 & 10:30:32.31 & +05:28:07.35 & 23.99 & 2.1113  & 2 & ELG & 2.15$_{-0.39}^{+0.40}$ & 1.84$_{-0.34}^{+0.34}$ & Ly$\alpha$ (3783\,\AA), NV\\
XID 098 & 10:30:31.05 & +05:27:47.32 & 27.3  & --  & 0 & None & 1.49$_{-0.57}^{+0.59}$  & 5.21$_{-0.89}^{+0.91}$ & None \\
XID 102 & 10:30:31.61 & +05:29:04.44 & 25.6  & 1.1084  & 2 & ELG &  5.44$_{-0.46}^{+0.46}$  & 1.72$_{-1.40}^{+1.78}$ & MgII, [OII], [NeV] \\
XID 103 & 10:30:29.79 & +05:29:09.09 & 24.65 & 1.4990  & 1 & ELG &  1.44$_{-0.20}^{+0.20}$ & 1.50$_{-0.22}^{+0.22}$  & [OII] \\
XID 153 & 10:30:28.29 & +05:27:40.72 & 27.02 & --  & 0 & None   & 2.16$_{-0.26}^{+3.26}$ & 4.44$_{-1.74}^{+1.74}$  & None \\
XID 198 & 10:30:39.95 & +05:26:55.23 & 24.1  & --  & 0 & None &  1.35$_{-0.17}^{+0.15}$  & 1.02$_{-0.22}^{+0.26}$  &  None \\
XID 206 & 10:30:14.76 & +05:29:19.80 & $>$27.5 & 1.124  & 1 & ELG &  1.48$_{-1.00}^{+3.44}$ & 1.48$_{-1.00}^{+3.44}$  &  [OII], [NeV]   \\
XID 214 & 10:30:44.59 & +05:25:44.77 & 25.97 & --  & 0 & None  & 1.38$_{-0.01}^{+0.54}$  & 1.44$_{-0.01}^{+0.92}$ & None \\
XID 365 & 10:30:27.88 & +05:23:48.31 & 25.72 & 1.4159  & 1 & ELG &  1.83$_{-0.37}^{+0.39}$ & 1.73$_{-0.31}^{+0.29}$  &  [OII], [OIII] \\
XID 366 & 10:30:28.58 & +05:32:24.61 & 24.52 & 1.0474 & 2 & ELG & 0.92$_{-0.22}^{+0.22}$ & 0.84$_{-0.20}^{+0.20}$ &   [OII], [NeIII]\\ 
XID 369 & 10:30:31.28 & +05:28:02.30 & 24.7  & 0.8248 & 2 & ELG  & 1.17$_{-0.07}^{+0.05}$   & 1.17$_{-0.03}^{+0.05}$ &  [OII], [OIII], H$\gamma$, H$\beta$ \\
\hline
\hline
	\vspace{0.02cm}
\end{tabular}
}
\caption{\normalsize Summary of the redshift measurements obtained in the LBT-MODS campaign presented in this work. The identification number is taken from the \cha\ J1030 catalog \citep{nanni20}. $r_{\rm AB}$ is the AUTO magnitude in the $r$ band, as reported in \citet{nanni20}. $z_{\rm spec}$ is the source spectroscopic redshift, $Q_{\rm z}$ is the quality flag for $z_{\rm spec}$ (2: secure; 1: uncertain; 0: no measurement), ``Class'' is the spectral classification (ELG: emission line galaxy; None; no classification). $z_{\rm phot,M21}$ is the photometric redshift computed in \citet{marchesi21} and used to select the high-redshift candidates, while $z_{\rm phot,new}$ is the updated photometric redshift computed using the new $g-$ and Ks--band photometric information presented in Section~\ref{sec:new_data}, as described in Appendix~\ref{sec:logn-logs}.
}
\label{tab:redshift_summary}
\end{table*}
\endgroup

As reported in Table~\ref{tab:redshift_summary}, all the targets for which we computed a spectroscopic redshift in this work (both the primary targets and the fillers) are classified as emission line galaxies (ELGs). In M21, we defined ELGs as sources whose spectra do not show any detectable AGN features, but a flat or bluish continuum and low--ionization emission lines that are compatible with a typical star formation activity. These sources optical emission is therefore dominated by non-AGN processes, and we can assess their AGN nature only thanks to the X-ray detection, since their X-ray luminosity L$_{\rm 2-10keV}$ is way above the 10$^{42}$\,erg\,s$^{-1}$ threshold which is commonly adopted to select AGN in the X-rays.
Notably, in M21 the ELG population we sampled had much lower average redshift ($\langle z_{\rm ELG,M21}\rangle$=1.08, with standard deviation $\sigma_{\rm z,ELG,M21}$=0.58) and was therefore significantly brighter ($\langle r_{\rm AB,ELG,M21}\rangle$=23.0, with standard deviation $\sigma_{\rm r,ELG,M21}$=1.8; as a reference, the targets analyzed in this work have $\langle r_{\rm AB,ELG}\rangle$=25.3, with standard deviation $\sigma_{\rm r,ELG}$=1.0). In two targets (XID 102, $z$=1.1084; XID 206, $z$=1.124; see Figures~\ref{fig:spec_MODS_a} and \ref{fig:spec_MODS_b} in the Appendix) we find tentative evidence of the [NeV] narrow emission line, which is a known tracer of obscured AGN \citep[see, e.g.,][]{maiolino98,vignali06,vignali10,mignoli13,cleri22, barchiesi23}.

\begin{figure}[htbp]
 \centering
\includegraphics[width=0.98\linewidth,trim={0 0 1.3cm 1.3cm},clip]{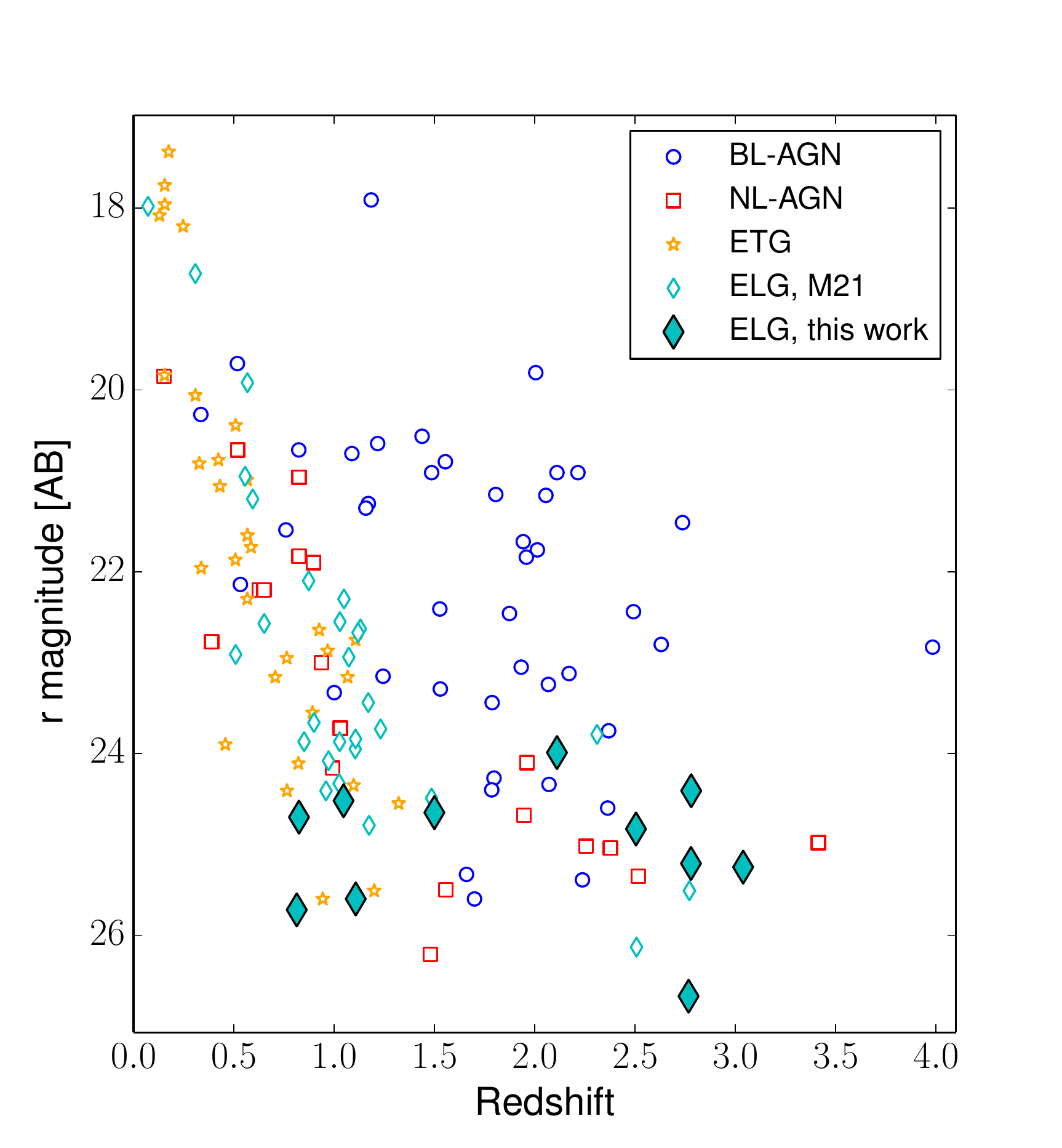}
\caption{\normalsize 
Total AB magnitude in the $r$ band from \citet{nanni20} as a function of spectroscopic redshift for the 122 \cha\ J1030 sources with $z_{\rm spec}$ from M21 (open markers): we do not include in this plot the $z$=6.3 QSO SDSSJ1030+0525, which is not detected in the $r$-band. BL-AGN are plotted as blue circles, narrow-line AGN as red squares, emission line galaxies as cyan diamonds, and early type galaxies as orange stars. The emission line galaxies reported in this work are plotted as full cyan diamonds.
}\label{fig:z_vs_r-mag_w_spec_type}
\end{figure}

From Table~\ref{tab:redshift_summary}, as well as from Figure~\ref{fig:z_vs_r-mag_w_spec_type}, where we report the \cha\ J1030 sources total AB magnitudes \citep[from][]{nanni20} as a function of spectroscopic redshift, it can also be understood how this spectroscopic campaign significantly increased the spectroscopic completeness of the faint ($r_{\rm AB}\geq$24) \cha\ J1030 population, from 28 to 40 sources.
It is worth noting that this new campaign, targeting fainter sources than in M21, led to the detection of no new broad line AGN (BL-AGN). In the M21 sample, 43 out of 123 sources with a spectroscopic redshift (35\,\%) were BL-AGN; as it can be seen in Figure~\ref{fig:z_vs_r-mag_w_spec_type}, BL-AGN are more easily detected at brighter magnitudes, so that 21 out of 50 sources with $r_{\rm AB}\leq$23 are BL-AGN (42\,\%), while out of 31 sources in M21 with $z_{\rm spec}$ and $r_{\rm AB}\geq$24 only 8 (26\,\%) were BL-AGN. Following the spectroscopic campaign reported in this work, the BL-AGN fraction at magnitudes $r_{\rm AB}\geq$24 decreased to 19\,\%.

In M21, we also extensively discussed how the lack of narrow-band photometry in the J1030 field makes BL-AGN the class of targets for which the photometric redshifts are significantly less accurate: more in detail, only 28 out of 43 BL-AGN (i.e., 65\,\%) had a photometric redshift in agreement with the spectroscopic one (where the agreement is achieved when ||$z_{\rm phot}$-$z_{\rm spec}$||/(1+$z_{\rm spec}$)$<$0.15). As a reference, 25 out of 28 ELGs (i.e., 89\,\%) had photometric redshifts in agreement with the spectroscopic ones. 
Consequently, the fact that we did not find any new BL-AGN in the \cha\ J1030 optically faint subsample has also the indirect result of strengthening the reliability of the photometric redshifts for the 108 targets lacking a $z_{\rm spec}$. In fact, the vast majority of them (87 out of 108, 81\,\%) have  $r_{\rm AB}\geq$24.

Finally,two of the targets in the sample (XID 008, $z_{\rm spec}$ = 2.78; XID 011, $z_{\rm spec}$ = 2.767) were also part of the sample studied in \citet{peca21}. That work analyzed a subsample of 54 \cha\ J1030 sources having hardness ratio (i.e., the ratio between the difference between 2--7\,keV and 0.5--2\,keV source counts, and 0.5--7\,keV source counts) HR$>-$0.1 and at least 50 net counts in the 0.5--7\,keV band. Since low-energy, soft photons are more easily absorbed by the obscuring material surrounding accreting supermassive black holes, sources with larger HR values are more likely to be obscured. In such targets, features like the fluorescent Iron K$\alpha$ line at 6.4\,keV and the 7.1\,keV Iron edge are particularly prominent and can be used to determine the source redshift through X-ray spectral fitting \citep[see, e.g.,][]{maccacaro04,civano05,vignali15,simmonds18,marchesi19b,iwasawa20}.
The X-ray spectrum of each of the 54 targets was then fit leaving the redshift free to vary, and thus using the X-ray features mentioned above to determine the targets redshifts. 

\citet{peca21} measured a X-ray redshift $z_{\rm X}$ = 2.80$_{-0.05}^{+0.05}$ for XID 008, where the Iron K$\alpha$ line was detected. Such a measurement is in excellent agreement with the spectroscopic redshift, particularly considering the uncertainties on the measurement, that are much smaller than the photometric redshift ones. The method was instead less effective in XID 011, where the X-ray redshift measurement ($z_{\rm X}$ = 1.94$_{-0.38}^{+1.05}$) was based on the tentative detection of the Iron edge. Even in this second target, however, the X-ray redshift is consistent with the spectroscopic one within the errors.

\begin{figure}[htbp]
 \centering 
\includegraphics[width=1.\linewidth]{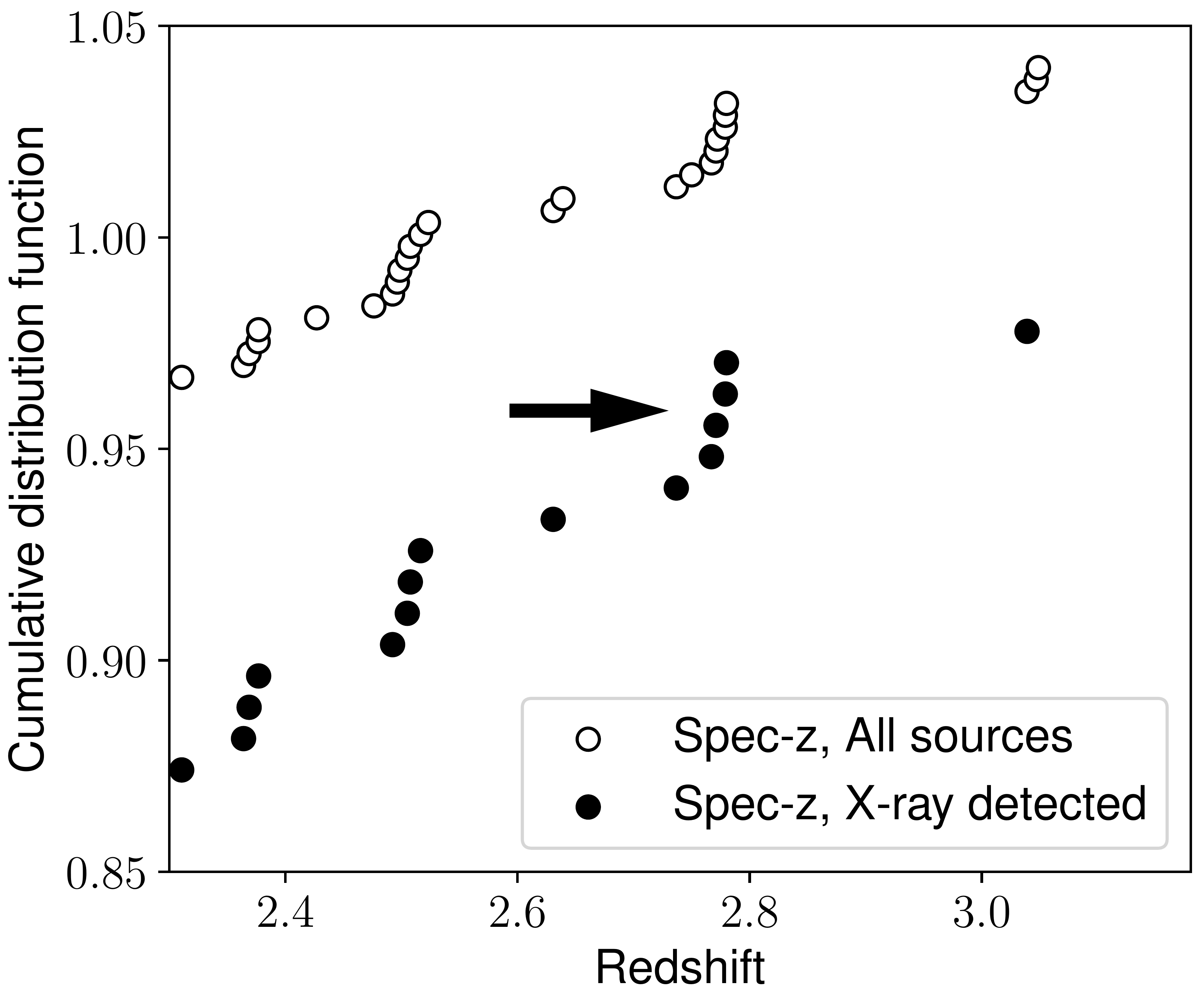}
\caption{\normalsize 
Redshift cumulative distribution function around $z$=2.78 for the \cha\ J1030 sources with a spectroscopic redshift (full markers), and for all the J1030 sources with a spectroscopic redshift, including the non-X-ray detected ones (empty markers). The overall cumulative distribution function is shifted by 0.1 for visualization purposes. The $z\sim$2.78 structure is marked with an arrow. Other two potential structures can be observed at $z\sim$2.37 and $z\sim$2.51 (see the text for more details).
}\label{fig:cdf}
\end{figure}

\section{A new structure at \texorpdfstring{$z\sim$2.8}{z~2.8}}\label{sec:z_28_LSS}
As mentioned in the introduction, faint X-ray selected AGN have been shown to be excellent tracers of large scale structures (LSSs), and in M21 we reported several J1030 structures detected in the \cha\ J1030 up to redshift $z\sim$1.5. Thanks to the LBT-MODS spectroscopic campaign presented in this work, we are now able to identify a new large scale structure in the J1030 field, this one located at a significantly larger redshift, $z\sim$2.78. We report in Table~\ref{tab:LSS} the overdensity members and their properties, while in Figure~\ref{fig:cdf} we plot the redshift cumulative distribution function around $z\sim$2.78 of both the \cha\ J1030 sources and of all the sources in J1030 having a spectroscopic redshift. As we already discussed in M21, such a representation provides a good visualization of candidate redshift structures in the field, which readily appear as sharp cumulative distribution function rises, and is free of binning issues. Four sources in the overdensity are X-ray detected, namely XID 007, XID 008, XID 011 and XID 109. In this work, we measured a spectroscopic redshift for XID 007, XID 008 and XID 011, while the redshift of XID 109 ($z$=2.771) was already reported in M21 and was also obtained using the MODS spectrograph. From this same plot, it can be noticed that two other potential overdensities are observed at $z\sim$2.37 and $z\sim$2.51. While the signal-to-noise ratios of these two group of sources are lower than the one we adopt to select an overdensity, S/N=3.8, we plan to further investigate these candidate structures in a follow-up work.

Besides the four X-ray $z\sim$2.78 sources, there are two more J1030 objects that are not X-ray detected and have $z_{\rm spec}\sim$2.78: for both targets, we measured the redshift using the Multi Unit Spectroscopic Explorer (MUSE) instrument mounted on the European Southern Observatory (ESO) Very Large Telescope (VLT). A $\sim$1$^\prime$$\times$1$^\prime$ region of the J1030 field centered on the $z$=6.3 quasar was observed by MUSE \citep{bacon10,bacon15} between June and July 2016 under the program ID 095.A-0714 for a total of $\sim$\,6.4\,hours of exposure time. The data reduction of the MUSE observations was presented in \citet{gilli19}. The final data cube has a wavelength range of 4750$-$9350\,\AA, a spectral sampling of 1.25\,\AA\ and a spatial sampling of 0.2$^{\prime\prime}$, covering an area of 1~sq-arcmin centered at the sky position of the QSO SDSSJ1030+0524. The 1D spectra were extracted by combining the spaxels inside a 3-pixel aperture that matches the seeing full width at half maximum of 0.6$^{\prime\prime}$. The MUSE wavelength coverage did not allow to access the Ly$\alpha$ region at $z\sim$2.78, but the high S/N spectra of the two galaxy members present a rich plethora of spectral features, both in emission (HeII~1640\,\AA \ and CIII]~1909\,\AA\ lines) and in absorption (ISM lines), that provided accurate redshift measurements via cross-correlation with a high-z star-forming template \citep{talia12}.

We report the flux--calibrated, one-dimensional spectra of the six members of the LSS in Figure~\ref{fig:spec_LSS}. The spectra of the remaining targets for which we measured a redshift are reported in Appendix~\ref{sec:app_spectra}.

\begin{figure*}[ht]
 \centering 
\includegraphics[width=1.\linewidth,trim={0cm 0cm 0cm 0cm},clip]{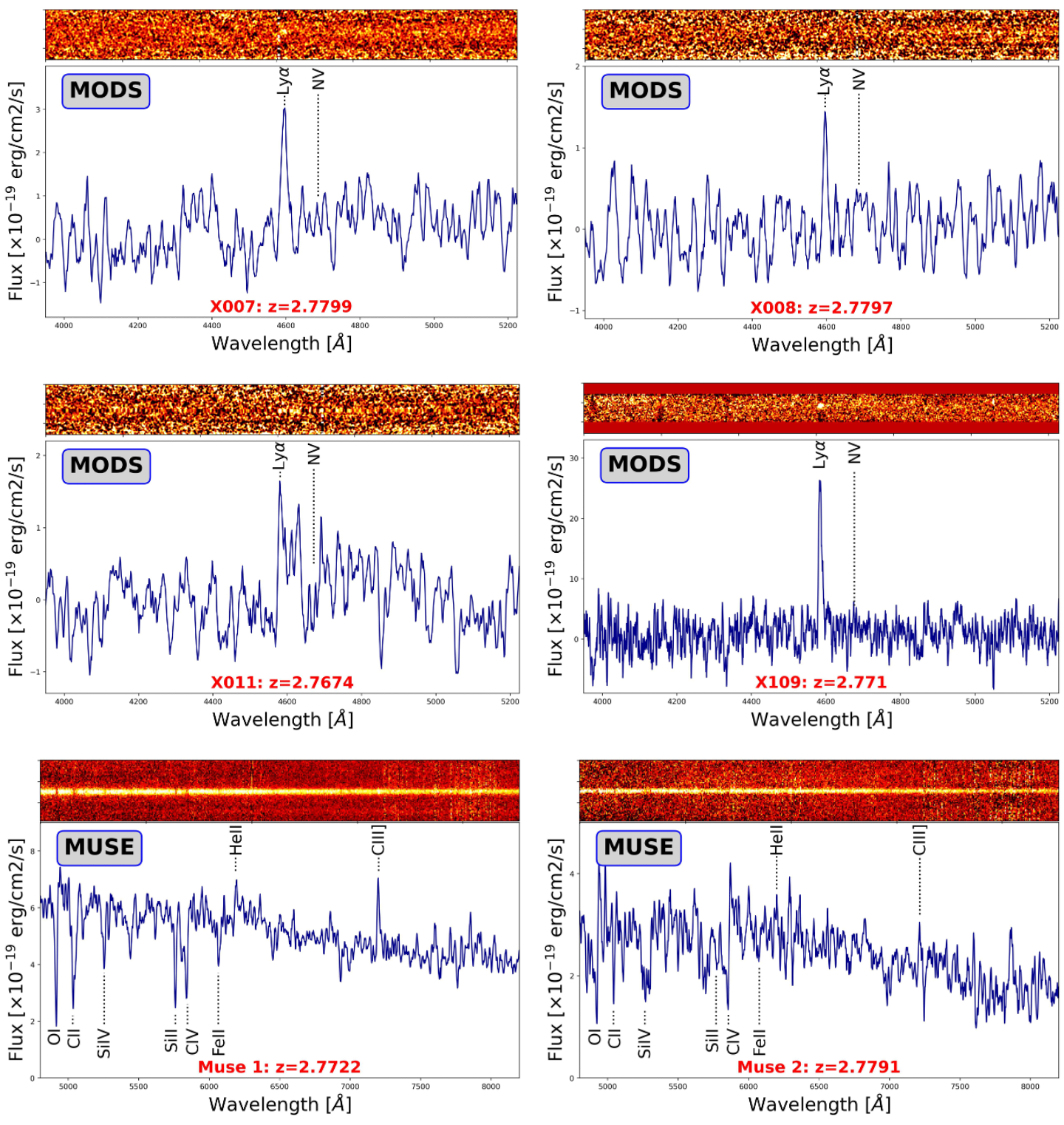}
\caption{\normalsize 
Flux-calibrated spectra of the six members of the $z\sim$2.78 large scale structure: in the top and central panels we report the LBT-MODS spectra of the four \cha\ sources, while in the bottom panels we show the VLT-MUSE spectra of the two sources that are not detected in the X-rays. The spectrum of XID 109 was obtained in the previous \cha\ J1030 spectroscopic campaign, which we presented in \citet{marchesi21}. We highlighted the expected positions of the main emission lines that fall in the observed spectral range. On top of each one-dimensional spectrum, we report the two-dimensional one.
}\label{fig:spec_LSS}
\end{figure*}

\begin{figure*}[htbp]
 \begin{minipage}[b]{.49\textwidth} 
 \centering 
 \includegraphics[width=0.98\linewidth]{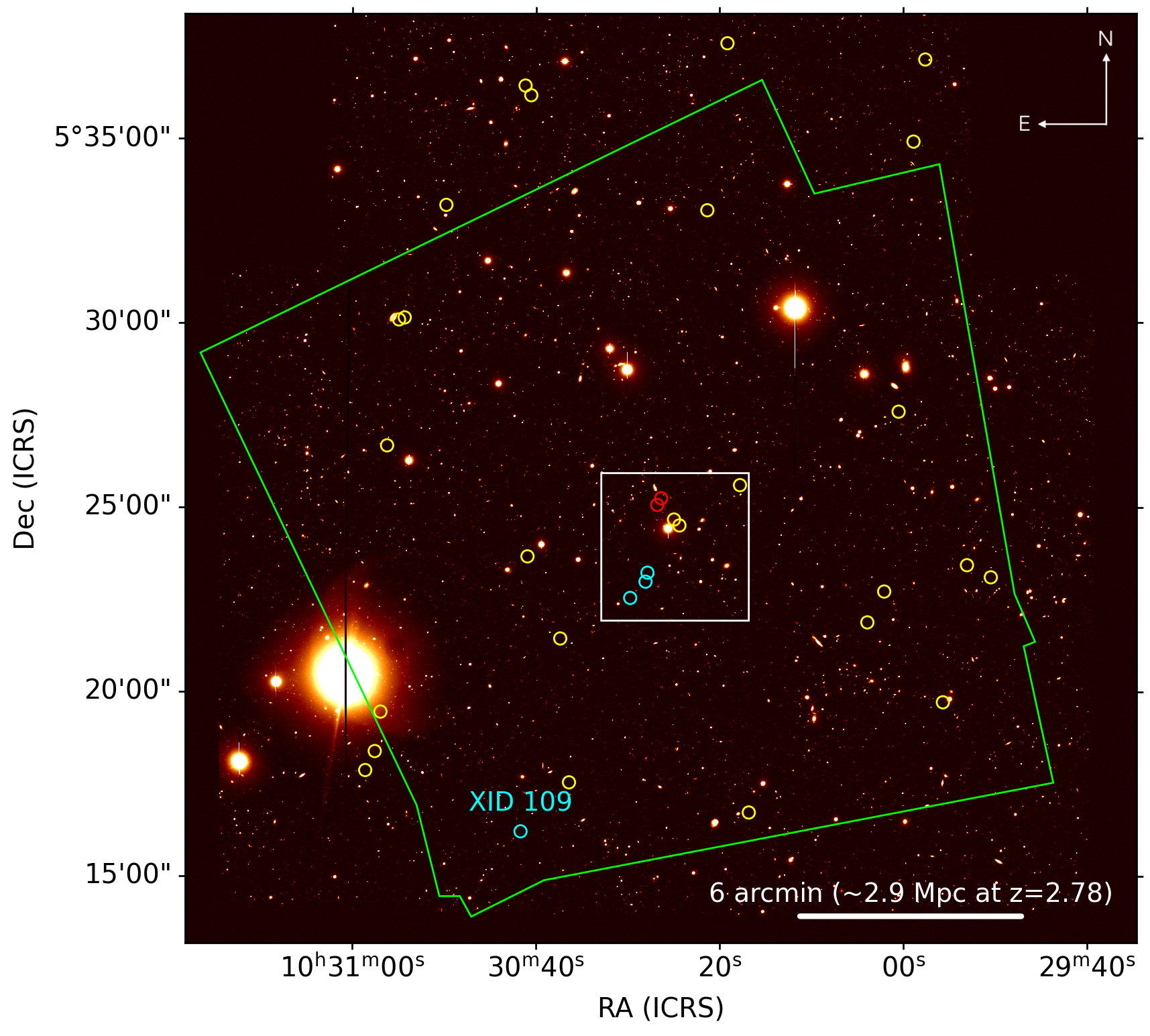}
 \end{minipage}
 \begin{minipage}[b]{.48\textwidth} 
 \centering 
 \includegraphics[width=0.98\linewidth]{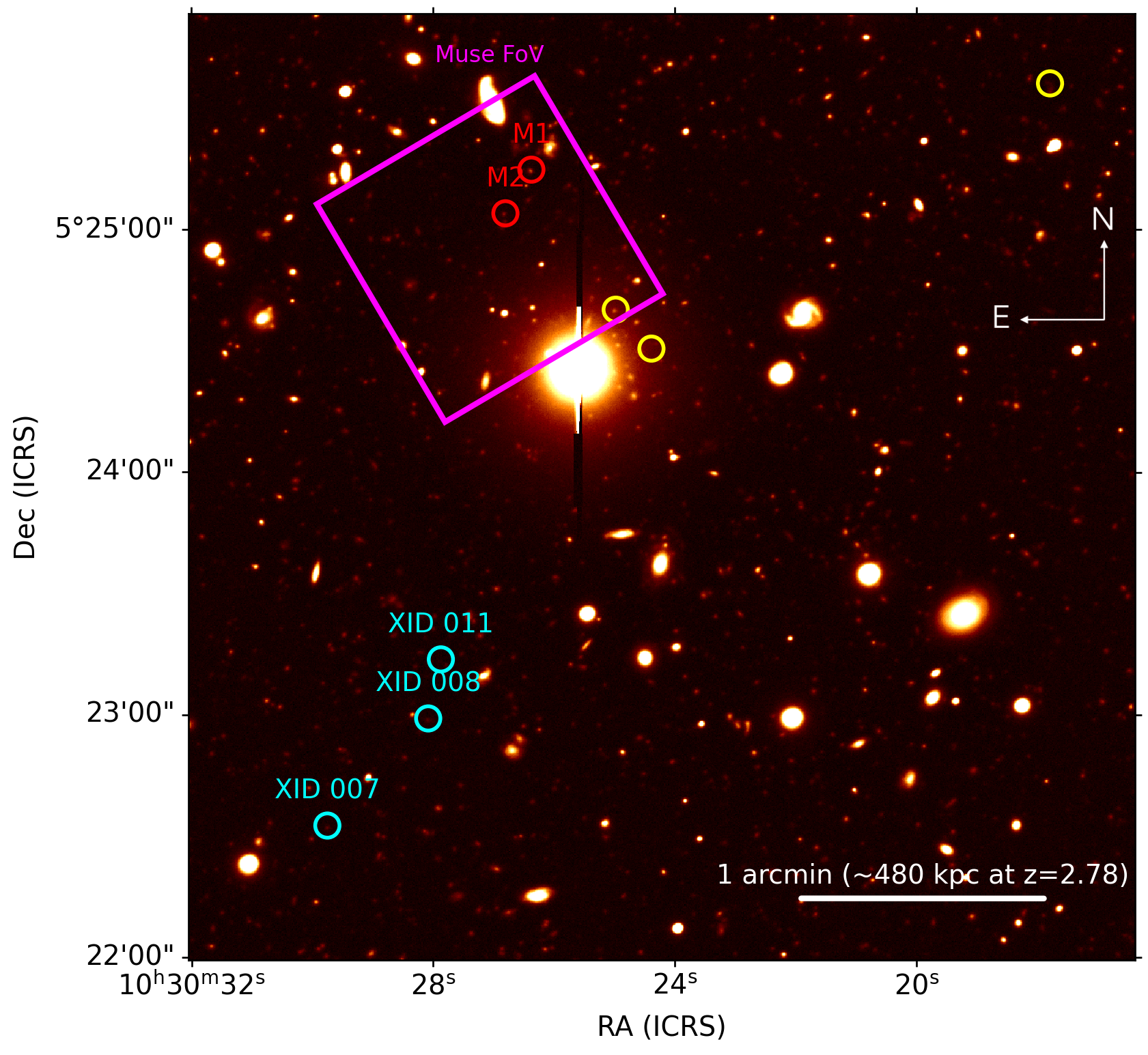}
 \end{minipage}
\caption{\normalsize LBT-LBC r-band images of the J1030 field. Left:
whole field image, with the \cha\ coverage shown in green. The spectroscopically confirmed, X-ray detected members of the $z\sim$2.78 large scale structure are plotted with a red circle, while the two non-X-ray detected sources with a MUSE spectroscopic redshift are plotted in red. Non-X-ray sources with a photometric redshift in the range $z_{\rm phot}$=[2.68--2.88] are plotted as yellow circles. 
Right: inset of the $\sim$4$^{\prime}\times$4$^{\prime}$ region (shown as a white box in the left panel) where we detect five out of six members of the LSS. The MUSE coverage of the J1030 field is also shown here for reference.
}\label{fig:LSS}
\end{figure*}

For consistency with the approach we used in M21, we use the method presented in \citet{gilli03}, and earlier in \citet{cohen99}, to compute the significance of the structure. Sources are first distributed in velocity space $V= c\,\rm{ln}(1+z)$ rather than in redshift space, since $dV$ corresponds to local velocity variations relative to the Hubble expansion. The velocity distribution is then smoothed twice: once with a broad Gaussian, having $\sigma_B$ = 1.5$\times$ 10$^4$\,km\,s$^{-1}$, to create the expected background distribution of the sample; the second one with a narrower Gaussian, having $\sigma_S$ = 300\,km\,s$^{-1}$, which is the typical velocity dispersion observed in galaxy structures \citep[e.g.,][]{cohen99,gilli03}. We then compute the signal-to-noise ratio S/N = $(S-B)/\sqrt{B}$, where $S$ is the number of sources within $\Delta V= \pm800$\,km\,s$^{-1}$ around the peak $z$=2.775 ($S$=4 in the X-ray-detected sample, $S$=6 in the whole one), and $B$ is the number of background sources in the same interval; $\Delta V$ was chosen to optimize the S/N value. We obtain a signal-to-noise ratio S/N=6.3 for the X-ray overdensity and S/N=5.5 when the non X-ray detected J1030 sources are included in the computation. The Poisson probability of observing $S$ sources given the estimated background value $B$ is P$_{\rm Poiss(N\geq N_{\rm obj,X})}$ = 3 $\times$ 10$^{-4}$ for the X-ray LSS, while we obtain P$_{\rm Poiss(N\geq N_{\rm obj,All})}$ = 5 $\times$ 10$^{-5}$ for the whole LSS. We note that this is among the highest redshift X-ray LSSs ever detected: for example, in the CDF-S 7\,Ms \citep{luo17} the two farthest X-ray detected LSSs are at $z\sim$2.57 and $z\sim$3.47. The SSA22 \cha\ survey \citep{lehmer09,razdom22} found instead 13 X-ray members of the SSA22 protocluster at $z$=3.09, which was however first detected through a optical spectroscopy follow-up of candidate $z\sim$3 Lyman break galaxies \citep{steidel98}. We note that the main difference between high-z LSSs detected in the optical with respect to those detected in the X-rays is the type of sources used to trace the LSS: in optical surveys the most common tracer are Lyman Break Galaxies \citep[see, e.g.,][]{toshikawa16} and Ly$\alpha$ and H$\alpha$ emitters \citep[e.g.,][]{higuchi19,koyama21}, while in the X-rays the tracers are AGN. Thus, by detecting and studying X-ray LSSs one can simultaneously study the properties of the host and those of the accreting SMBH, and compare the properties of the AGN with those of AGN at similar redshifts that are not located in overdensities.

We also compute the distance between the different structure members, assuming a flat $\Lambda$CDM cosmology (see Introduction). At the average redshift of the targets in the structure, $z$=2.775, the angular-to-physical conversion factor is 8.026\,kpc/$^{\prime\prime}$. Five out of six targets (namely, XID007, XID008, XID011 and the two MUSE detected galaxies) are separated by distances in the range $d_{\rm A}$=100--1360\,kpc (see Figure~\ref{fig:LSS}, right panel). The closest pairs of objects are the two MUSE--detected targets ($d_{\rm A}$=101\,kpc) and XID008 and XID011 ($d_{\rm A}$=120\,kpc). As shown in Figure~\ref{fig:LSS}, left panel, XID109 is located farther away from the rest of the objects, at $\sim$3.3--4.6\,Mpc. The difference in redshift between the targets can also be converted in angular diameter distances: for the sources in the LSS, it varies in the range 30\,kpc (XID 007 and XID 008; $\Delta z$=2$\times$10$^{-4}$) to 2\,Mpc (XID008 and XID011, $\Delta z$=1.25$\times$10$^{-2}$). We note that a separation of several Mpc between AGN belonging to the same structure was also previously reported in \citet{shen21}, where three AGN were found to be part of a protocluster at $z\sim$3.3. Two of these AGN lied within $\sim$1\,Mpc, while the third one was found at $\sim$5\,Mpc from the remaining two. 

\begingroup
\renewcommand*{\arraystretch}{1.5}
\begin{table*}
\centering
\scalebox{0.76}{
\vspace{.1cm}
 \begin{tabular}{cccccccccc}
 \hline 
 \hline
ObjID   &    RA      &    DEC      & $z$ &  $N_{\rm H}$ & log(L$_{\rm 2-10keV}$) & log(M$_*$/M$_{\odot}$) & SFR$_{\rm SED}$ & SFR$_{\rm 1.4GHz}$ & log(L$_{\rm 1.4GHz}$)\\ 
        & hh:mm:ss   & dd:mm:ss    &        & 10$^{22}$\,cm$^{-2}$   & erg s$^{-1}$ & & M$_{\odot}$ yr$^{-1}$ & M$_{\odot}$ yr$^{-1}$ & W Hz$^{-1}$\\
 \hline
XID 007 & 10:30:29.8 & +05:22:32.3 & 2.7799  & $<$3.9 & 44.08$_{-0.08}^{+0.08}$ & 10.49$\pm$0.24 & 165$\pm$107 & 103$\pm$35 & 23.52$_{-0.18}^{+0.13}$ \\
XID 008 & 10:30:28.1 & +05:22:58.8 & 2.7797  & 47.0$_{-21.7}^{+23.1}$ & 44.46$_{-0.12}^{+0.11}$ & 10.36$\pm$0.29 & $<$188  & $<$96  & $<$23.47 \\
XID 011 & 10:30:27.9 & +05:23:13.4 & 2.7674  & 39.4$_{-18.6}^{+20.0}$ & 44.90$_{-0.12}^{+0.11}$ & 10.24$\pm$0.71 & 82$\pm$74  & $<$91 & $<$23.58 \\
XID 109 & 10:30:41.7 & +05:16:12.5 & 2.771  & 2.0$_{-2.0}^{+8.3}$ & 44.48$_{-0.10}^{+0.11}$ & 11.10$\pm$0.13 & 234$\pm$188 & 197$\pm$61 & 23.89$_{-0.16}^{+0.12}$ \\ 
\hline
MUSE 1  & 10:30:26.4 & +05:25:14.5 & 2.7722& N/A & $<$43.15  & 9.49$\pm$0.48  & 85$\pm$81 & $<$165 & $<$23.58 \\
MUSE 2  & 10:30:26.8 & +05:25:03.7 & 2.7791 & N/A & $<$43.15 & 9.11$\pm$0.69  & $<$87 & $<$226 & $<$23.66 \\
\hline
Photometric candidates & -- & -- & 2.78$\pm$0.10 & N/A & N/A & 9.97 (0.43) & 38.3 (60.9) & -- & -- \\
\hline
\hline
	\vspace{0.02cm}
\end{tabular}
}
\caption{\normalsize Properties of the members of the $z\sim$2.78 large scale structure detected in this work. $z$ is the source redshift. $N_{\rm H}$ and L$_{\rm 2-10keV}$ are the source column density and intrinsic, 2--10\,keV luminosity as computed from the X-ray spectrum \citep{signorini23} Since the two MUSE targets are not detected in the X-rays, their $N_{\rm H}$ cannot be computed: the L$_{\rm 2-10keV}$ upper limit is computed from the \cha\ J1030 survey flux limit in the 2--10\,keV band ($f_{\rm 2-10keV,lim}$=2.7$\times$10$^{-16}$\,erg\,s$^{-1}$\,cm$^{-2}$). M$_*$ is the stellar mass content of the host galaxy (in units of solar masses) and SFR$_{\rm SED}$ is the host star formation rate: both these quantities have been computed through spectral energy distribution fitting. We also report, for comparison, the median and standard deviation of the stellar mass and star formation rates of the 26 galaxies in the J1030 field whose photometric redshift is consistent with $z$=2.78 (see the text for more details). Finally, SFR$_{\rm 1.4GHz}$ is the SFR independently derived from the 1.4\,GHz luminosity L$_{\rm 1.4GHz}$ measured from the JVLA J1030 survey \citep{damato22}. The SFR$_{\rm 1.4GHz}$ values reported here are obtained assuming a Chabrier initial mass function. Upper limits are reported at the 3\,$\sigma$ confidence level.
}
\label{tab:LSS}
\end{table*}
\endgroup

\subsection{Candidate overdensity members based on their photometric redshift}\label{sec:photo_LSS}
By combining the new $g-$ and $K_s-$band imaging data presented in Section~\ref{sec:new_data} with the existing multi-band photometry, we computed photometric redshifts for all the $K_s-$selected sources in the J1030 field (M. Mignoli et al. in prep.). We report in Table~\ref{tab:photometry_summary} the optical/near-infrared photometric bands we use to compute these photometric redshifts. Following the same approach we used in M21, the spectral energy distributions (SEDs) of the sources are fitted using two different codes: \texttt{Hyperz} \citep{bolzonella00} and \texttt{EAzY} \citep{brammer08}. We adopt the \citet{calzetti00} extinction law to take reddening into account, and we use 75 templates dominated by stellar emission, following the work by \citet{ilbert13}. More in detail, 19 out of 75 are empirical templates from the SWIRE template library \citep{polletta07}:  7 for elliptical galaxies, 12 for different classes of spiral galaxies (S0, Sa, Sb, Sc, Sd and Sdm). Other 12 templates describe the SED of starburst galaxies, although no emission lines are included in the template. Finally, the remaining  44 templates are based on the \citet{bruzual03} stellar population synthesis models, and have been first introduced in \citet{ilbert09,ilbert13}.

We use photometric redshifts to identify potential members of the $z$=2.78 overdensity. To do so, we conservatively select only those sources which have photometric redshift in the range $z_{\rm phot}$=[2.68--2.88] according to both \texttt{Hyperz} and \texttt{EAzY}. We choose such a range because the RMS of the difference between spectroscopic and photometric redshifts in the J1030 sample at $z>$2 is $\Delta z\sim$\,0.1. Overall, our sample of $z\sim$2.78 photometric candidates contains 26 galaxies. We report in Figure~\ref{fig:LSS} the distribution of the sources over the J1030 field.

\subsection{Host properties of the overdensity members}
To better understand the properties of the members of the $z\sim$2.78 overdensity, we take advantage of the excellent multi-wavelength coverage of the J1030 field to perform a self-consistent SED fitting from the X-rays to the near infrared. To do so, we use the CIGALE v2022.1 tool \citep{yang20,yang22}. CIGALE v2022.1 fits a source SED and computes the host galaxy physical properties. With respect to the CIGALE tool \citep{burgarella05,boquien19}, CIGALE v2022.1 allows one to also fit the X-ray information, if available. This update to the original tool is particularly helpful to put stronger constraints on the AGN contribution to the SED and, consequently, to obtain more accurate measurements of the host properties.

For the X-ray detected sources and photometric redshift candidates, we use the optical/near-infrared photometric information reported in Table~\ref{tab:photometry_summary}. For the four X-ray-detected AGN we also use the 0.5--2\,keV and 2--7\,keV fluxes reported in \citet{nanni20}. Finally, both MUSE targets are detected in our new K-band image of the J1030 field. We therefore performed aperture photometry at the K$_S$ position of the sources in the $griz$ LBT/LBC bands, as well as in the WIRCAM/CFHT Y and J ones. The two MUSE targets were also detected in the MUSYC BVR catalog, and the MUSE 1 source was also detected in the K-Deep catalog. We therefore included in our SEDs the MUSYC U and V magnitudes from the BVR catalog and the MUSYC H magnitude from the K Deep one. We note that there is an excellent agreement ($\Delta {\rm mag}\leq$0.05) between our B and K magnitudes and the MUSYC B$_{\rm BVR}$ and K$_{\rm Kd}$, a result that support our choice of using the additional MUSYC data points like we did for the X-ray sources. We also searched the NASA/IPAC Infrared Science Archive\footnote{\url{https://irsa.ipac.caltech.edu/}} for a WISE counterpart of the two MUSE targets, but we did not find any.

\begingroup
\renewcommand*{\arraystretch}{1.5}
\begin{table*}
\centering
\scalebox{0.95}{
\vspace{.1cm}
 \begin{tabular}{ccccc}
 \hline
 \hline
 Catalog & Filters & Area & 5\,$\sigma$ Depth & Reference \\
     &     &   & m$_{\rm AB}$\\
  \hline    
  MUSYC BVR & UBV & 30$^\prime\times$30$^\prime$ & 26.0, 26.2, 26.0  & \citet{gawiser06}  \\
  LBT-LBC & $g$ & 23$^\prime\times$25$^\prime$ & 26.4 & This work. \\
  LBT-LBC & $riz$ & 23$^\prime\times$25$^\prime$ & 26.8, 26.1, 25.6 & \citet{morselli14}\\
  HST-ACS & F775W & 3.3$^\prime\times$3.3$^\prime$ & 27.5 & \citet{stiavelli05} \\
  HST-ACS & F850LP & 3.3$^\prime\times$3.3$^\prime$ & 27.5 & \citet{kim09} \\
  WIRCAM-CFHT & YJ & 24$^\prime\times$24$^\prime$ & 24.7, 24.4 & \citet{balmaverde17}\\
  HST-WFC3 & F160W & 2$^\prime\times$2$^\prime$ & 27.5 & \citet{damato20}\\
  MUSYC K Deep & H & 10$^\prime\times$10$^\prime$ & 23.6 & \citet{quadri07}\\
  WIRCAM-CFHT & $K_s$ & 24$^\prime\times$24$^\prime$ & 23.4 & This work. \\
  \textit{Spitzer}-IRAC & CH1-2 & 35$^\prime\times$35$^\prime$ & 22.7, 22.4 & \citet{annunziatella18}\\
  \textit{Spitzer}-IRAC & CH3-4 & 22$^\prime\times$15$^\prime$ & 22.1, 21.8 & IRSA Archive\\
  \hline
	\hline
\end{tabular}}
	\caption{\normalsize Properties of the photometric catalogs used in this work.
	}
\label{tab:photometry_summary}
\end{table*}
\endgroup

In Table~\ref{tab:LSS} we report the stellar mass M$_*$ and star formation rate (SFR) of the six $z\sim$2.78 LSS members, as well as the median and standard deviation of both parameters for the sample of 26 galaxies with a photometric redshift consistent with $z$=2.78. As it can be seen, the four X-ray detected targets are significantly more massive than the two MUSE-detected ones: the stellar mass of the X-ray targets varies in the range log(M$_*$/M$_{\odot}$)=[10.2--11.1], while the two non-X-ray detected objects both have log(M$_*$/M$_{\odot}$)$<$10. The sample of candidate overdensity members having photometric redshift $z_{\rm phot}$=[2.68--2.88] is also, on average, less massive than the X-ray detected sources. More specifically, the photometric sample has median stellar mass log(M$_*$/M$_{\odot}$)=9.97 (with standard deviation $\sigma$=0.43); thus, all X-ray sources have stellar masses above the median mass of the photometric population. We also note that our photometric galaxies sample is $K-$selected, and $K-$selected samples are biased in favor of more massive hosts \citep[see, e.g.,][]{dunne09}. This further strengthens the significance of detecting X-ray sources in more massive hosts.

As it can be seen in Figure~\ref{fig:spec_LSS}, the two MUSE targets are brighter (in the B band which broadly matches the wavelength range shown in the spectra) than the X-ray targets. More specifically, the two MUSE targets have $B{\rm _AB}\sim$25, while the X-ray targets all have $B{\rm _AB}>$26, and actually all but one even have $B{\rm _AB}>$27. Things change, however, when moving to redder wavelengths. In particular, the X-ray targets are significantly brighter in the K-band ($K_{\rm AB}$=[22.2–-22.7]) than the MUSE ones ($K_{\rm AB}$=[23.4--24.8]). Such a different behavior at different wavelengths is directly linked to the different host masses observed in the two subsamples: the MUSE targets are likely young, compact sources with intense star formation processes ongoing, while the X-ray AGN are hosted in older (thus redder), more massive galaxies.

Massive galaxies, and in particular those sources with log(M$_*$/M$_{\odot}$)$>$10, have been shown to be among the best tracers of large scale structures and more specifically of filaments in the cosmic web \citep[see, e.g.,][]{malavasi17,sarron19,kuchner20}. In our case, such a tentative evidence clearly needs to be validated with a larger, statistically  more reliable sample. Nonetheless, the significant mass difference between the X-ray and the non-X-ray detected host galaxies supports a scenario where the X-ray AGN (i.e., those sources with X-ray luminosity L$_{\rm 2-10keV}>$10$^{42}$\,erg\,s$^{-1}$) are hosted by the most massive galaxies in a structure and are therefore more likely to be powered by more massive and more efficient accreting supermassive black holes. If such a result will be confirmed in larger samples, it would (at least partially) explain why X-ray detected targets are such efficient tracers of LSSs.

Quantitatively, using the \citet{suh20} M$_{\rm BH}$--M$_*$ relation, we find that the X-ray detected targets have black hole masses in the range  M$_{\rm BH}$  = [6$\times$10$^{6}$\,M$_\odot$--7$\times$10$^{7}$\,M$_\odot$]. Since the \citet{suh20} relation was calibrated using X-ray emitting sources and galaxies with log(M$_*$/M$_{\odot}$)$>$10, it cannot be reliably used to estimate the black hole mass of the two MUSE-detected sources. Interestingly, the most massive object among the members of the structure is XID 109, the only one which is not in spatial proximity with the others. Since more massive galaxies are expected to be located at the center of the large scale structure \citep[this is, for example, the case for the protocluster detected at redshift $z\sim$1.7 in the J1030 field; see][]{gilli19,damato20}, such a result could hint at the fact that the center of the structure does not coincide with the region where we detected five out of six LSS members. Indeed, as shown in Figure~\ref{fig:LSS}, left panel, no clear evidence of clustering is observed when including in the overdensity the photometric redshift candidates. However, any further consideration can be validated only by a spectroscopic follow-up campaign of these candidates.

\begin{figure}[htbp]
 \centering 
\includegraphics[width=1.\linewidth,trim={0 0 1cm 2cm},clip]{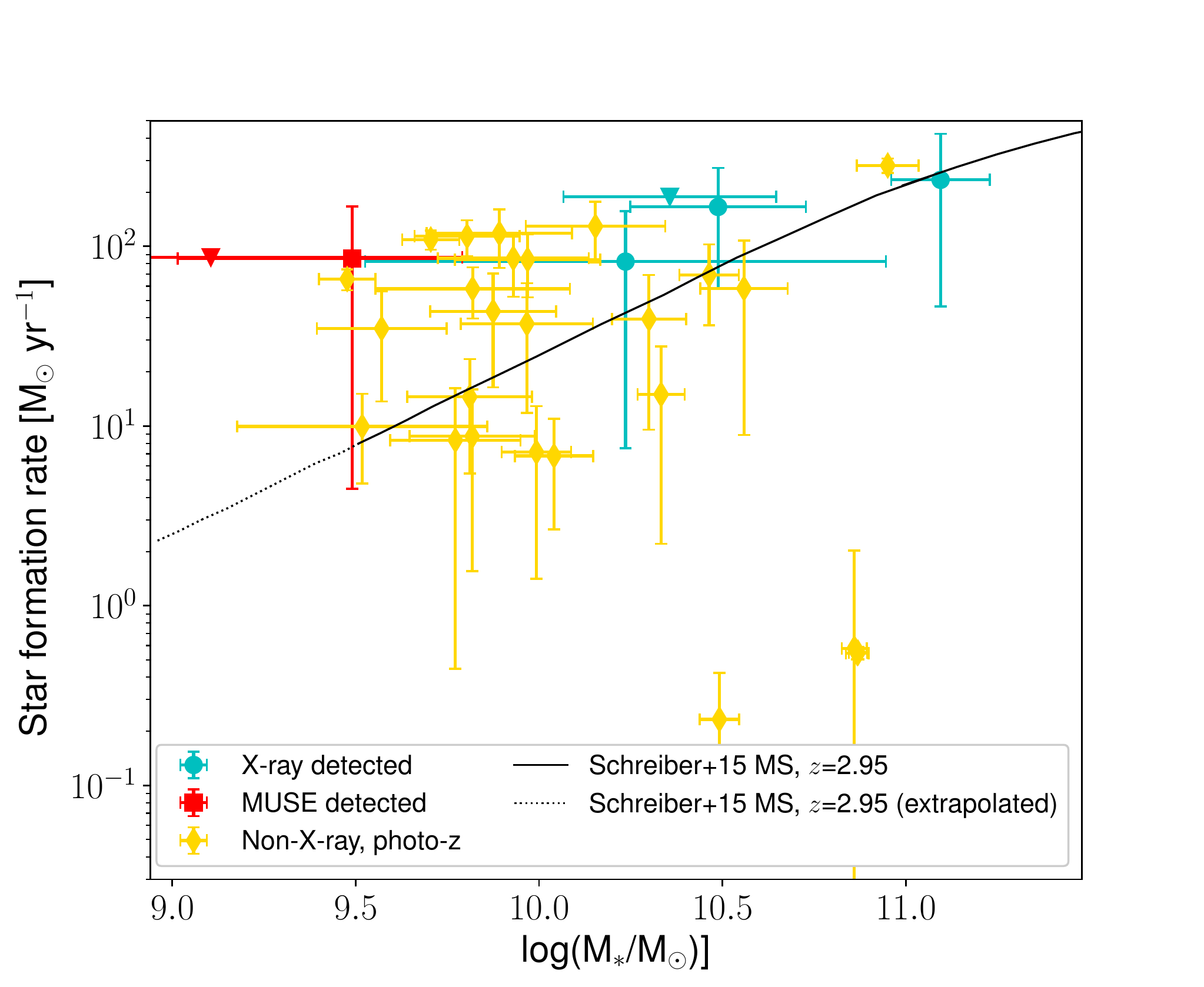}
\caption{\normalsize 
Star formation rate as a function of stellar mass for the four X-ray detected (cyan circles) and the two MUSE detected (red squares) members of the $z\sim$2.78 overdensity, as well as for the 26 candidate members of the overdensity having a photometric redshift $z_{\rm phot}$=[2.68--2.88] (yellow diamonds). Upper limits are plotted as downwards triangles. As a reference, we also plot the $z$=2.95 main sequence from \citet{schreiber15}. Both M$_*$ and SFR are computed through SED fitting. 
}\label{fig:SFR_mstar}
\end{figure}

We plot in Figure~\ref{fig:SFR_mstar} the star formation rate as a function of the stellar mass for the six spectroscopically confirmed targets in the LSS, as well as for the 26 photometric candidates. As a reference, we also plot the \citet{schreiber15} M$_*$-SFR main sequence at $z\sim$2.95. As it can be seen, the X-ray detected sources tend to lie on the MS. One of the two MUSE targets has instead a nominal SFR value well above the MS one for a galaxy with stellar mass log(M$_*$/M$_{\odot}$)$\sim$9.5; however, the uncertainties on SFR for this target are large, and our measurement is consistent with the MS one within the 90\,\% confidence uncertainty. Finally, the majority of photometric candidates also tend to lie on the MS. A few high-mass sources, however, have SFR much lower than the expected one, thus suggesting that they could already be in a passive, non star-forming phase. This tentative trend, if spectroscopically confirmed, could then suggest that the X-ray sources in the LSS are hosted not only by the most massive galaxies, but also by objects that are still gas-rich and star-forming.

In Table~\ref{tab:LSS} we also report, for the four \cha--detected targets, the source column density as computed via X-ray spectral fitting \citep{signorini23}. Several observational works have shown evidence for an increasing fraction of obscured AGN (i.e., sources having column density $N_{\rm H}>$10$^{22}$\,cm$^{-2}$ moving from the Local Universe to redshift $z\sim$2--3 \citep[e.g.,][]{lafranca05,tozzi06,liu17,lanzuisi18,vito18,iwasawa20,peca23}, where sources with $N_{\rm H}>$10$^{23}$\,cm$^{-2}$ can reach 70--80\,\% of the overall AGN population. In such a scenario, nuclear obscuration from the so-called obscuring torus, that accounts for almost the totality of obscuration at $z\sim$0, must be complemented at higher redshifts by non-nuclear absorption taking place within the host galaxy. In particular, a recent work by \citet{gilli22} shows that the increasing density with redshift of the interstellar medium (ISM) can by itself explain the obscured fractions observed at $z\sim$2--3.

To test this scenario, we work under the assumption that galaxies where the ISM is denser are also those where the SFR is larger. We report in Table~\ref{tab:LSS} the SFR measurements obtained through SED fitting for the six members of the overdensity. Two out of four X-ray detected targets (XID 008 and XID 011) have $N_{\rm H}>$10$^{23}$\,cm$^{-2}$: in XID 011 we measure significant star formation activity (SFR$_{\rm SED}\sim$80\,M$_{\odot}$ yr$^{-1}$), while in XID 008 we can only compute a fairly loose upper limit (SFR$_{\rm SED}<$188\,M$_{\odot}$ yr$^{-1}$). The two remaining X-ray detected members of the overdensity (XID 007 and XID 109) both have a column density N$_{\rm H}<$10$^{23}$\,cm$^{-2}$: however, for both targets we measure a fairly large star formation rate, SFR$_{\rm SED}>$100\,M$_\odot$. However, the fact that we measure no significant obscuration in the X-ray spectrum of XID 007 and XID 109 could hint at the fact that the source SED is at least partially contaminated by the AGN emission, which would imply that the SFR derived from the SED might be over-estimated.

As a consistency check, we therefore independently computed the star formation rate of the four \cha\ J1030 sources in the structure by using the radio flux density obtained from a deep JVLA L-band observation of the J1030 field \citep{damato22}. XID 007 and 109 are both detected in the JVLA image, while for the remaining two targets we computed a 3\,$\sigma$ upper limit (see Table~\ref{tab:LSS}). Since \citet{damato22} reported that their fluxes $S_{\rm 1.34\,GHz}$ are measured at 1.34\,GHz, we applied a correction factor to compute the 1.4\,GHz luminosities reported in Table~\ref{tab:LSS}, as follows

\begin{equation}
    L_{\rm 1.4GHz} = \frac{4\pi D^2_{\rm L}(z)}{(1+z)^{(1+\alpha)}}\left(\frac{1.4\,{\rm GHz}}{1.34\,{\rm GHz}}\right)^\alpha S_{\rm 1.34\,GHz},
\end{equation}

where $\alpha$ is the radio photon index and is assumed to be $\alpha$=-0.7 for all targets.

We then used the equation reported in \citet{delvecchio21} to derive the SFR,

\begin{equation}
\frac{SFR}{M_\odot\,yr^{-1}} = f_{\rm IMF}\,\times\,10^{-24}\,10^{q_{\rm TIR}(z)}\,\frac{L_{\rm 1.4GHz}}{\rm W\,Hz^{-1}},
\end{equation}

where L$_{\rm 1.4GHz}$ it the 1.4\,GHz luminosity, while $f_{\rm IMF}$ is a correction factor related to the initial mass function (IMF) used in the computation: $f_{\rm IMF}$=1 for a Chabrier IMF; $f_{\rm IMF}$=1.7 for a Salpeter IMF. Finally $q_{\rm TIR}$($z$) is a factor that parameterizes the evolution with both redshift and the host stellar mass of the IR-radio correlation, 

\begin{multline}
   q_{\rm TIR}(z,M_*)\,=\,(2.646\pm0.024)\,\times\,(1+z)^{-0.023\pm0.008} \\ -({\rm log}M_*/M_{\odot}-10)\,\times\,(0.148\pm0.013).
\end{multline}

Since the stellar masses and star formation rates computed with CIGALE v2022.1 adopted a Chabrier IMF, in Table~\ref{tab:LSS} we report the radio SFR values obtained assuming $f_{\rm IMF}$=1. As it can be seen, there is a general good agreement between the SED fitting measurements and the 1.4\,GHz ones. In particular, for XID 007 and XID 109 we measure again SFR$_{\rm 1.4GHz}>$100\,M$_\odot$\,yr$^{-1}$. We remind, as a caveat, that the \citet{delvecchio21} relation has been computed using star-forming galaxies, rather than AGN, so the values obtained here are likely upper limits, as part of the radio emission in our targets may be of nuclear origin.

In summary, we do not observe a clear trend between the l.o.s. column density computed through X-ray spectral fitting and the host star formation rate computed either through SED fitting or through the 1.4\,GHz luminosity. Such a result is not in contradiction with the trends discussed in \citet{gilli22}, since those should be applied statistically on a large sample of sources, and not on single targets. For example, all the known $z\sim$6 quasars are by definition unobscured AGN \citep[see, e.g.,][]{nanni17,vito19} despite being hosted by young, gas-rich galaxies experiencing significant star formation activity.

\section{Summary and conclusions}\label{sec:summary}
In this work, we presented the results of a recent LBT-MODS spectroscopic follow-up campaign of a sample of \cha--detected sources in the J1030 field. We also used the LBT-LBC $g$ band and the CFHT-WIRCAM $K_{\rm s}$ band photometry we recently obtained on the J1030 field to update the photometric redshifts of the $z>$2 candidates in the \cha\ J1030 sample. The following are the main results of our analysis.

\begin{enumerate}
    \item Out of seven high-$z$ ($z_{\rm phot}>$2.7) candidates, we measured a spectroscopic redshift for five sources: all of them have $z_{\rm spec}>$2.5, thus proving the effectiveness of our photometric redshifts in selecting high-$z$ candidates.
    \item All the 12 sources for which we measured a new spectroscopic redshift are classified as emission line galaxies, which means that their optical emission is dominated by non-AGN processes. This result suggests that as we increase the sample spectroscopic completeness at faint magnitudes (all 12 targets have $r_{\rm AB}>$24) the contribution of the BL-AGN to the overall population significantly decreases. 
    \item The LBT-MODS spectroscopic campaign led to the serendipitous discovery of a large-scale structure at $z\sim$2.78 in the J1030 field. The structure contains six spectroscopically confirmed targets, four of which are X-ray detected. Five out of six targets are located within a $\sim$1$^\prime$ radius (i.e., within an angular distance $d_{\rm A}<$1.5\,Mpc). Other 26 galaxies have a photometric redshift in the range $z_{\rm phot}$=[2.68--2.88] and are therefore candidate members of the overdensity.
    \item We measure a large mass difference between the  X-ray detected AGN and the non-X-ray detected galaxies in the overdensity: the stellar mass of the X-ray targets varies in the range log(M$_*$/M$_{\odot}$)=[10.2--11.1], while the two spectroscopically confirmed, non-X-ray detected objects both have log(M$_*$/M$_{\odot}$)$<$10, and the photometric candidates have median stellar mass log(M$_*$/M$_{\odot}$)=10.0. Massive galaxies are known to be among the best tracers of large scale structures: this result, while limited by the small size of the sample, therefore provides a potential explanation of why X-ray detected AGN also are efficient LSSs tracers. 
    \item We measure in two independent ways -- through SED fitting and from the 1.4\,GHz luminosity -- the star formation rate of the galaxies hosting the $z\sim$2.78 overdensity. The two methods give consistent results, and the sources (particularly the X-ray detected ones) are found to lie on the \citet{schreiber15} SFR-M$_*$ main sequence.
    \item With this campaign, we measured a spectroscopic redshift for 12 \cha\ J1030 sources, thus increasing the sample spectroscopic completeness to 53\,\% (135 out of 256 sources).
\end{enumerate}

Future works will focus on obtaining photometric redshifts for the whole non-X-ray population (M. Mignoli et al. in prep.) in the J1030 field, and searching for both new members of the LSS and more in general for promising high-z candidates. With a larger number of sources with good photometric redshifts at $z\geq$2 it will be possible to search for high-redshift overdensities using statistical methods such as the Voronoi tessellation Monte Carlo algorithm \citep[see, e.g.,][]{cucciati18,lemaux18,shen21,lemaux22}. These photometric redshifts will also be used to select candidates for new spectroscopic follow-up campaigns with LBT-MODS and other facilities.

\section*{Acknowledgments}
We thank the referee for the detailed report and the useful suggestion. We acknowledge the support from the LBT-Italian Coordination Facility for the execution of observations, data distribution and reduction. The LBT is an international collaboration among institutions in the United States, Italy and Germany. LBT Corporation partners are the University of Arizona on behalf of the Arizona university system; Istituto Nazionale di Astrofisica, Italy; LBT Beteiligungsgesellschaft, Germany, representing the Max-Planck Society, the Astrophysical Institute Potsdam, and Heidelberg University; The Ohio State University, and The Research Corporation, on behalf of The University of Notre Dame, University of Minnesota and University of Virginia. We acknowledge financial contribution from the agreement ASI-INAF n. 2017-14-H.O. MB acknowledges financial support from the Italian L'Oreal UNESCO "For Women in Science" program and the PRIN MIUR 2017PH3WAT “Blackout”.

\bibliographystyle{aa}
\bibliography{Marchesi23_J1030_z278_LSS}

\onecolumn
\appendix

\section{Updated \cha\ J1030 photometric redshifts and refined high-redshift number counts}\label{sec:logn-logs}
As mentioned in Section~\ref{sec:new_data}, we recently observed the J1030 field in the $g$ and Ks band. We thus decided to re-run the \texttt{Hyperz} tool \citep{bolzonella00} to obtain more reliable photometric redshifts, particularly at z$\geq$2, where a deeper Ks coverage can significantly improve the overall quality of the spectral energy distribution (SED) fitting. The new photometric redshifts were computed by using the photometric bands reported in Table~\ref{tab:photometry_summary}. We followed the same SED fitting approach we used in M21, to which we refer for a complete description of the different steps. We also generated the redshift probability distribution function (PDZ) of each target.

\begin{figure}[htbp]
 \centering
\includegraphics[width=0.5\linewidth,trim={0 0 1.3cm 1.3cm},clip]{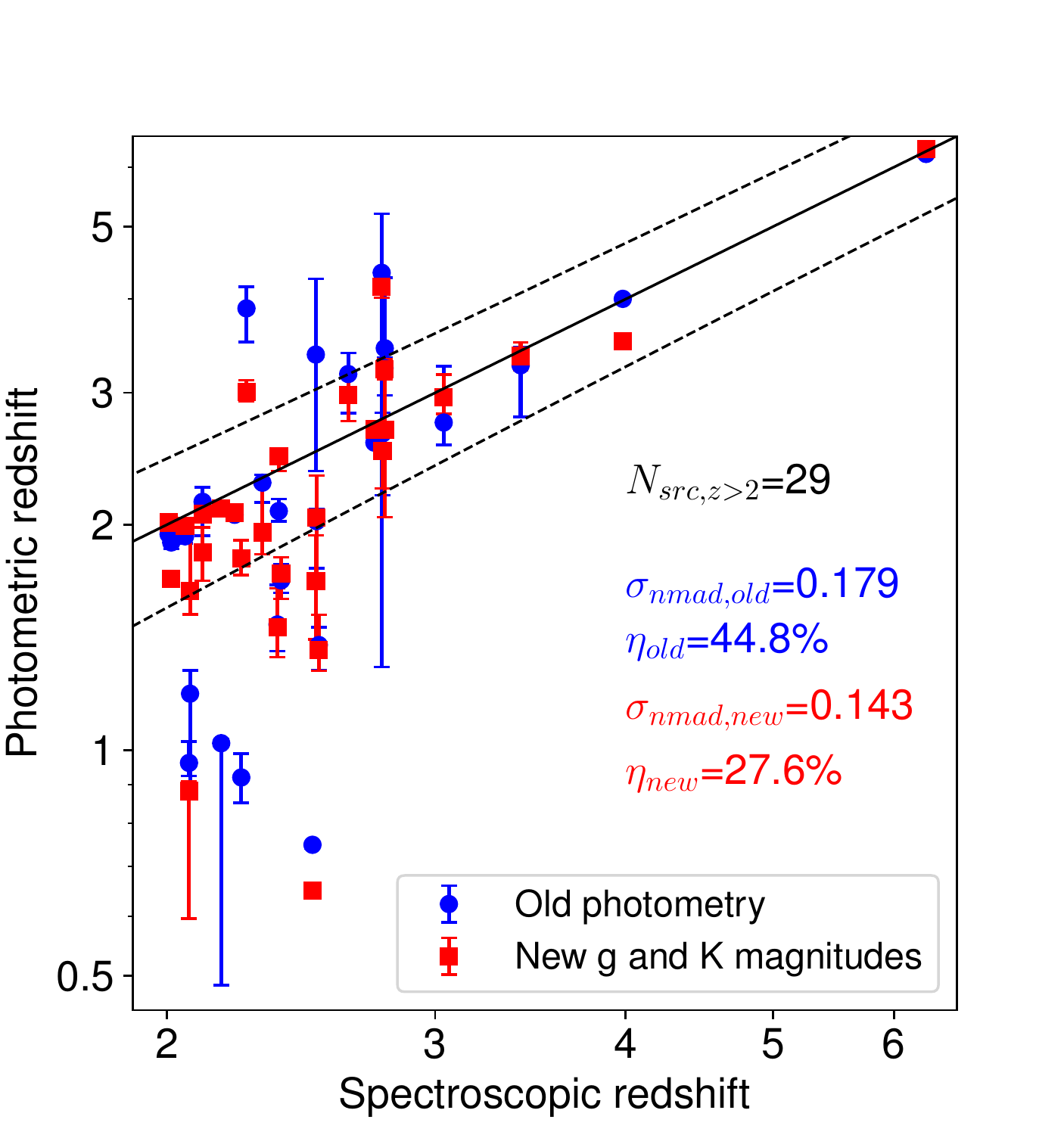}
\caption{\normalsize 
Photometric redshift as a function of the spectroscopic one for the 29 J1030 sources with spectroscopic redshift $z_{\rm spec}\geq$2. The black solid line marks the $z_{\rm spec}$ = $z_{\rm phot}$ relation, while the dashed lines contain the $z_{\rm phot}$ = $z_{\rm spec}\pm$0.15(1+$z_{\rm spec}$) region. Sources within this region are deemed to have a reliable photometric redshift. We note that all the outliers with $z_{\rm phot}<$1.2 are BL-AGN, a class of sources for which photometric redshifts are generally less reliable (see M21 for more details).
}\label{fig:spec_vs_phot}
\end{figure}

There are 29 \cha\ J1030 sources with $z_{\rm spec}\geq$2. We report in Figure~\ref{fig:spec_vs_phot} the distribution of the old and new photometric redshifts as a function of the spectroscopic ones. Following the standard approach to estimate a photometric redshift reliability we already used in M21, we assume that $z_{\rm spec}$ and $z_{\rm phot}$ are in agreement when ||$z_{\rm phot}$-$z_{\rm spec}$||/(1+$z_{\rm spec}$)$<$0.15. The new photometric redshifts are significantly more accurate than the old ones: 21 out of 29 objects (72\,\%) have $\Delta z$/(1+$z_{\rm spec}$)$<$0.15, with normalized median absolute deviation \citep{hoaglin83} $\sigma_{\rm NMAD,new}$ = 1.48 $\times$ median[||$z_{\rm phot,new}$-$z_{\rm spec}$||/(1+$z_{\rm spec}$)] = 0.143. In M21, instead, only 16 out of 29 targets (55\,\%) showed an agreement between the photometric redshift and the spectroscopic one, with  $\sigma_{\rm NMAD,M21}$=0.179. All the 8 outliers in the new photometric redshift sample are also outliers in the ``old'' photometric sample; furthermore, two of these outliers (XID011 and XID022) have $z_{\rm phot,new}>$3. Therefore, while in these two targets the photometric redshifts did not accurately pinpoint the actual source redshift, they were still effective in hinting at a high-redshift solution.

In M21 we computed the $z>$3 number counts in the \cha\ J1030 field. We found evidence of a potential overdensity at $z>$4 with respect to both other X-ray surveys and the predictions of different AGN population synthesis models. However, when we published these results only three out of 25 $z>$3 sources had a spectroscopic redshift, one of them (the only one at $z_{\rm spec}>$4) being the $z$=6.3 J1030 quasar itself. The $z>$4 overdensity we measured was therefore based on the photometric redshifts, and in particular on the sources redshift probability distribution functions (PDZs).

The spectroscopic campaign we presented in this work was therefore aimed, among other things, at increasing the \cha\ J1030 spectroscopic completeness at $z>$3, to put stronger constraints on the overdensity detection. As reported in Table~\ref{tab:redshift_summary}, all the high-$z$ sources which we were able to spectroscopically confirm are in the range $z_{\rm spec}$=[2.5--3.1]. This, combined with the new photometric redshifts, could in principle affect the \cha\ J1030 number counts measurements, and in particular reduce the high$-z$ excess presented in M21. For this reason, we recomputed the \cha\ J1030 $z>$3 number counts following the same approach presented in M21: most importantly, for sources lacking a spectroscopic redshift and having just a photometric redshift, we did not use the sources nominal best-fit redshift value, but rather their probability distribution functions. As discussed in M21, this is now a common approach when computing the high-$z$ number counts of X-ray surveys, since their spectroscopic completeness strongly decreases at $z>$2.5--3 \citep[see, e.g.,][]{marchesi16b,vito18}.

\begin{figure*} 
\begin{minipage}[b]{.32\textwidth} 
 \centering 
 \includegraphics[width=0.99\textwidth,trim={0 0 1.3cm 1.3cm},clip]{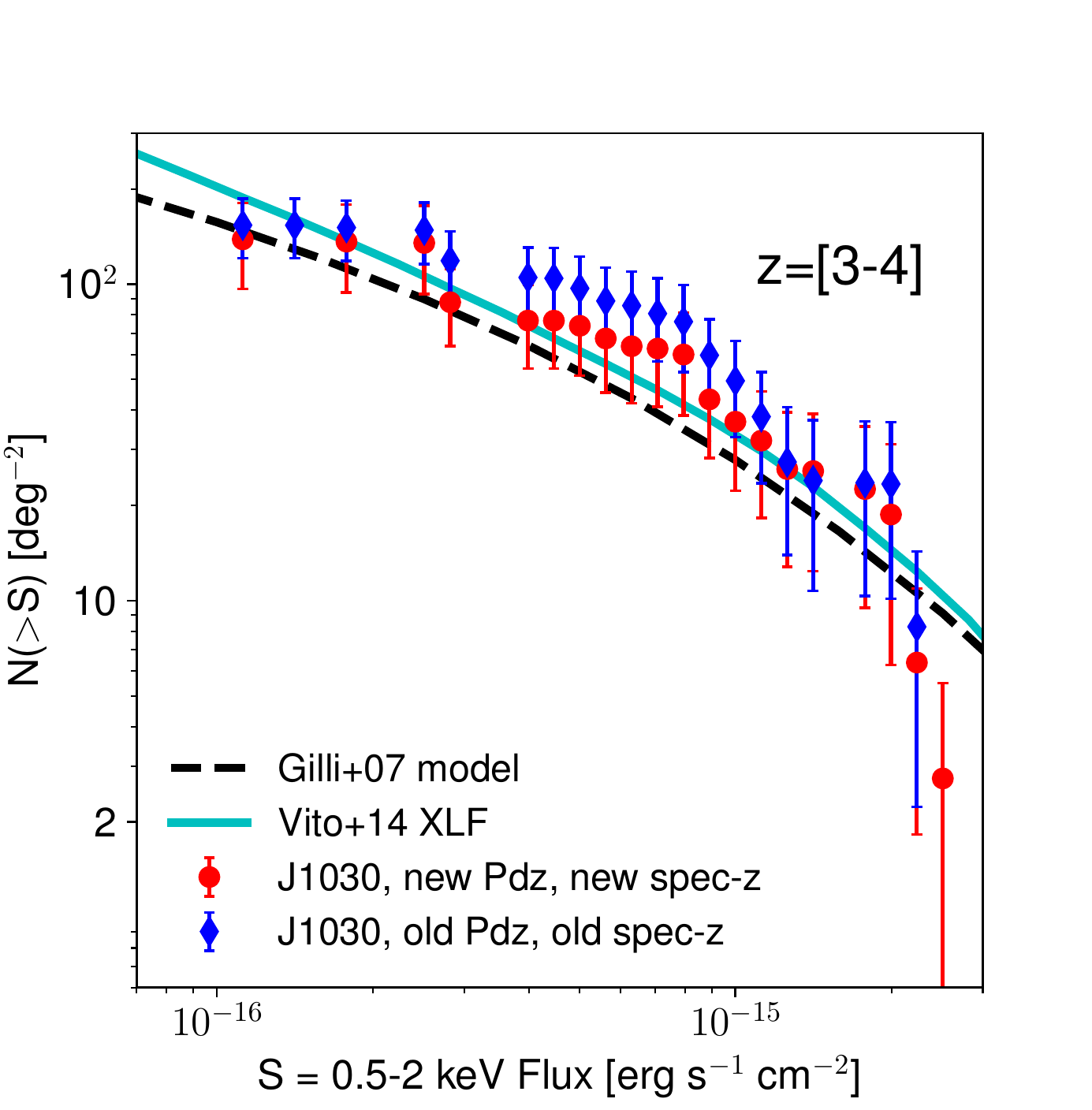} 
 \end{minipage} 
\begin{minipage}[b]{.32\textwidth}
 \centering 
 \includegraphics[width=0.99\textwidth,trim={0 0 1.3cm 1.3cm},clip]{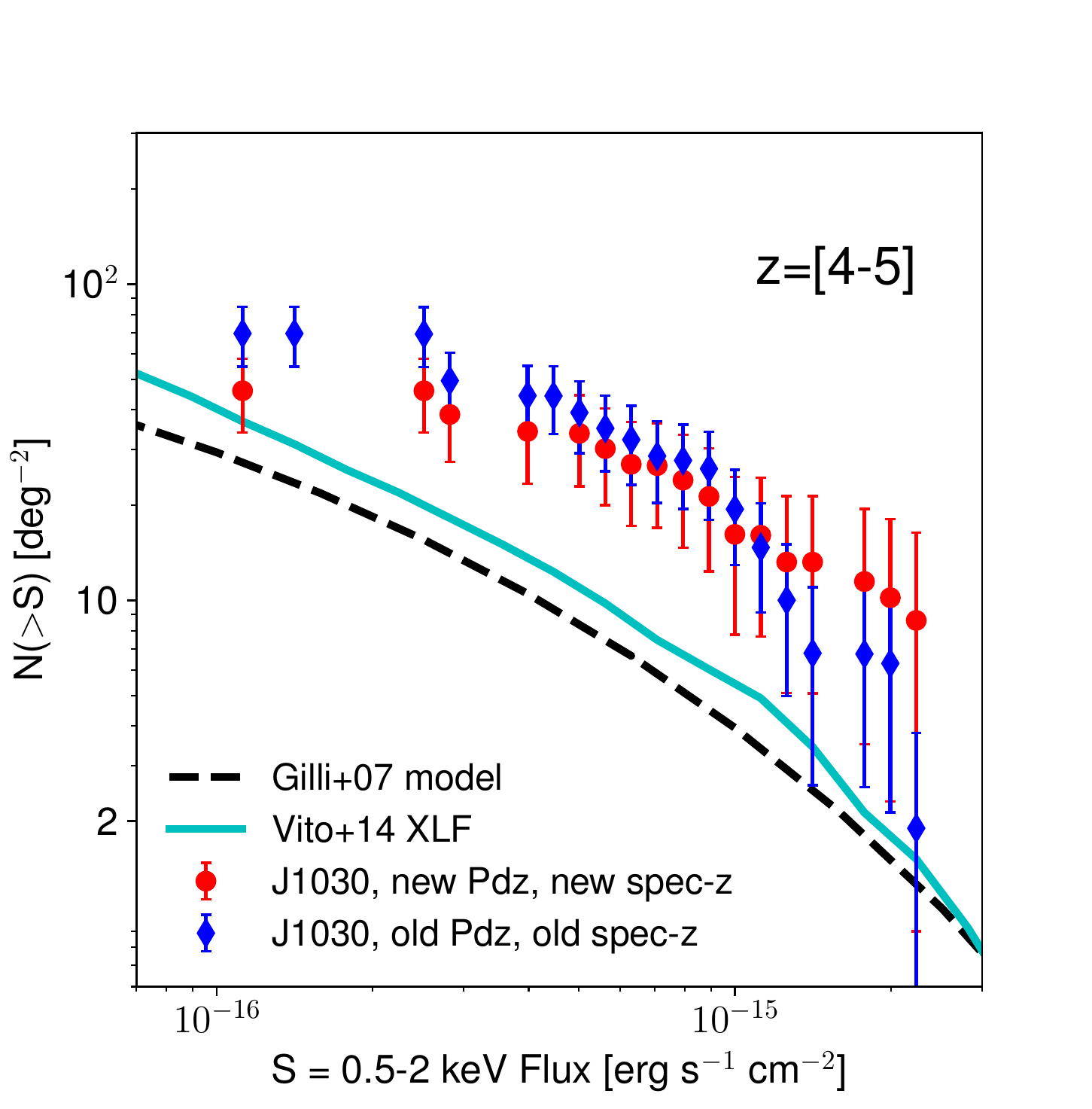} 
 \end{minipage} 
\begin{minipage}[b]{.32\textwidth}
 \centering 
 \includegraphics[width=0.99\textwidth,trim={0 0 1.3cm 1.3cm},clip]{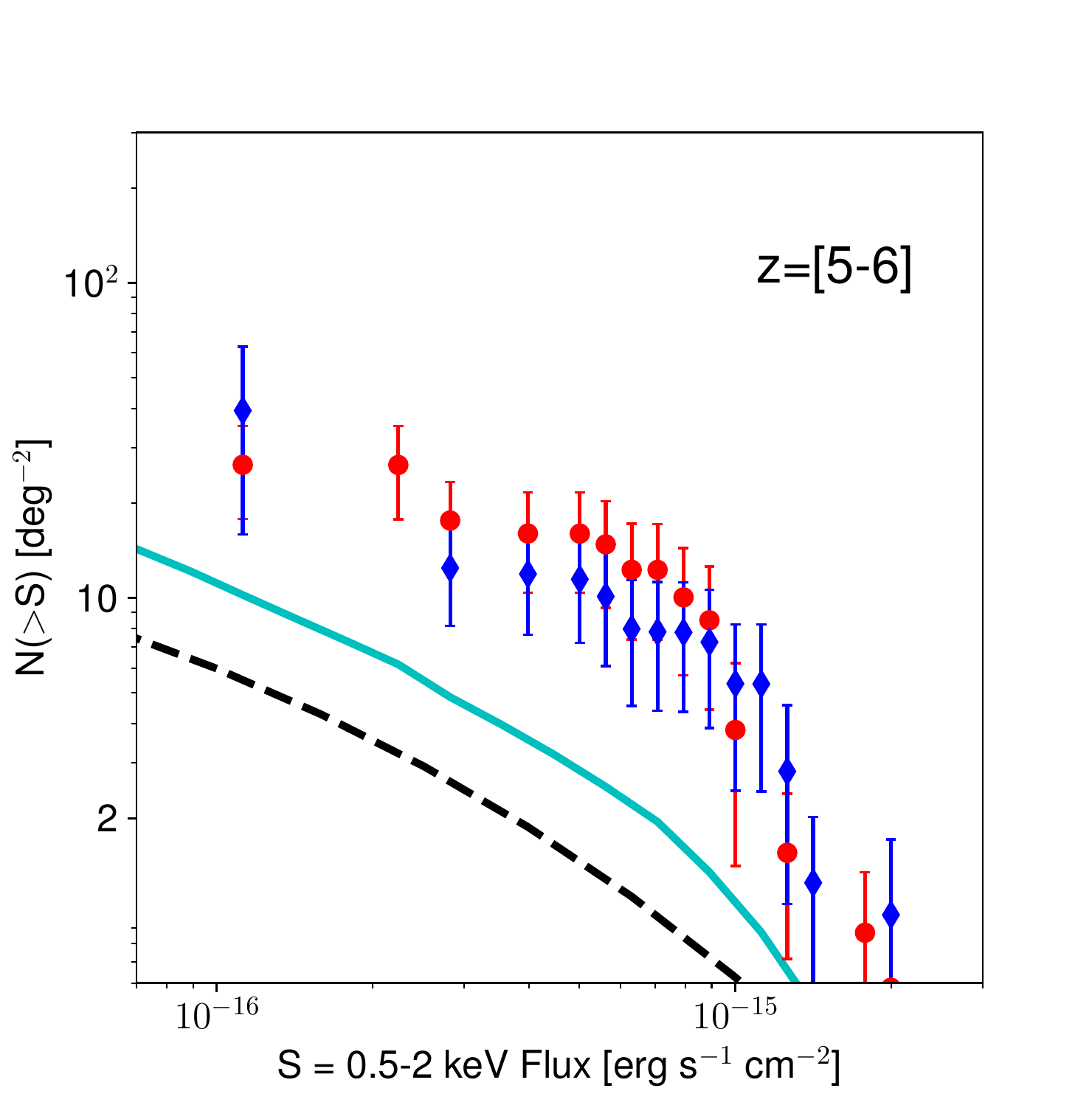} 
\end{minipage}
\caption{\normalsize \cha\ J1030 0.5--2\,keV number counts in the redshift bins $z$=[3--4] (left), $z$=[4--5] (center), and $z$=[5--6] (right). We plot as red circles the LogN-LogS obtained using both the new spectroscopic redshifts and the new photometric redshifts presented in this work; we instead plot as blue diamonds the M21 number counts. 
The number counts in the same redshift ranges derived using the \citet{gilli07} AGN population synthesis model (black dashed line) and the \citet{vito14} X-ray luminosity function (cyan solid line) are also shown for comparison. The overdensity measured in M21 at $z>$4 is still present in the newly refined number counts.
}\label{fig:logn-logs_z_bin}
\end{figure*}

We report in Figure~\ref{fig:logn-logs_z_bin} the \cha\ J1030 number counts in the redshift bins $z$=[3--4], $z$=[4--5] and $z$=[5--6]: we plot the original number counts reported in M21 as well as the number counts computed using both the new spectroscopic redshifts and the new photometric redshifts. For comparison, we also report the predictions of the \citet{gilli07} AGN population synthesis model and the number counts derived from the \citet{vito14} $z>$3 AGN X-ray luminosity function. As it can be seen, the difference between the number counts computed in M21 and those re-computed in this work using the new spectroscopic and photometric redshifts is marginal, and all number counts are consistent within the uncertainties. In particular, the excess observed at redshift $z>$4 in M21 is still present, and suggests that one or more additional overdensities can be present in the J1030 field between the $z$=2.78 and the $z$=6.3 one. We are now working on the computation of photometric redshifts for all the sources in the J1030 field (M. Mignoli et al. in prep.) and use them to search for more high-redshift candidates to then follow-up with spectroscopic campaigns like the one presented in this paper.

\section{Spectra}\label{sec:app_spectra}
\begin{figure*}[htbp]
 \centering 
\includegraphics[width=1.\linewidth,trim={0cm 7cm 0cm 0cm},clip]{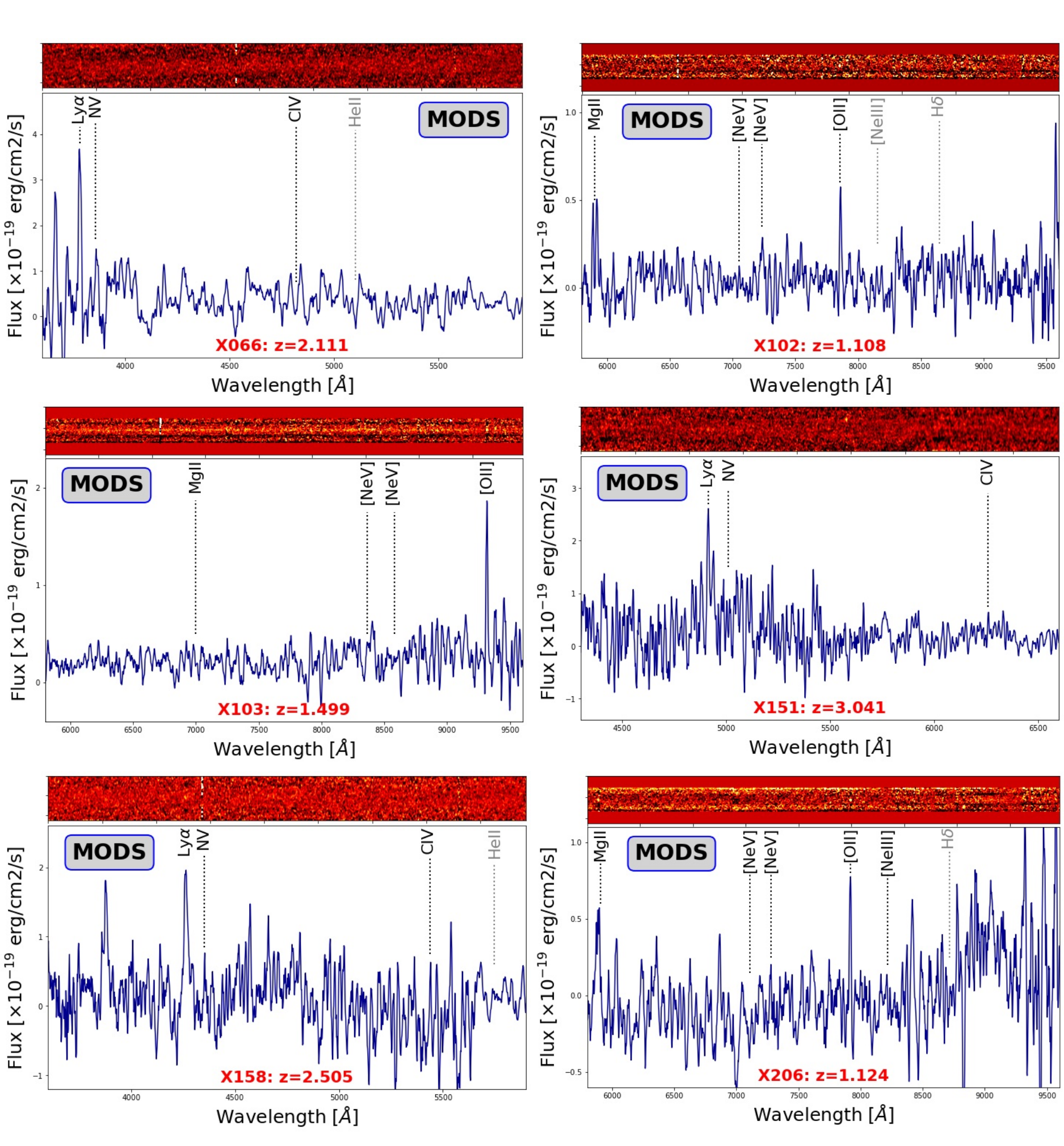}
\caption{\normalsize LBT-MODS
flux-calibrated spectra of six \cha\ J1030 sources targeted by the campaign presented in this work and not belonging to the $z\sim$2.78 structure. We highlighted the expected positions of the main emission lines that fall in the observed spectral range. On top of each one-dimensional spectrum, we report the two-dimensional one.
}\label{fig:spec_MODS_a}
\end{figure*}

\begin{figure*}[htbp]
 \centering 
\includegraphics[width=1.\linewidth,trim={0cm 13.5cm 0cm 0cm},clip]{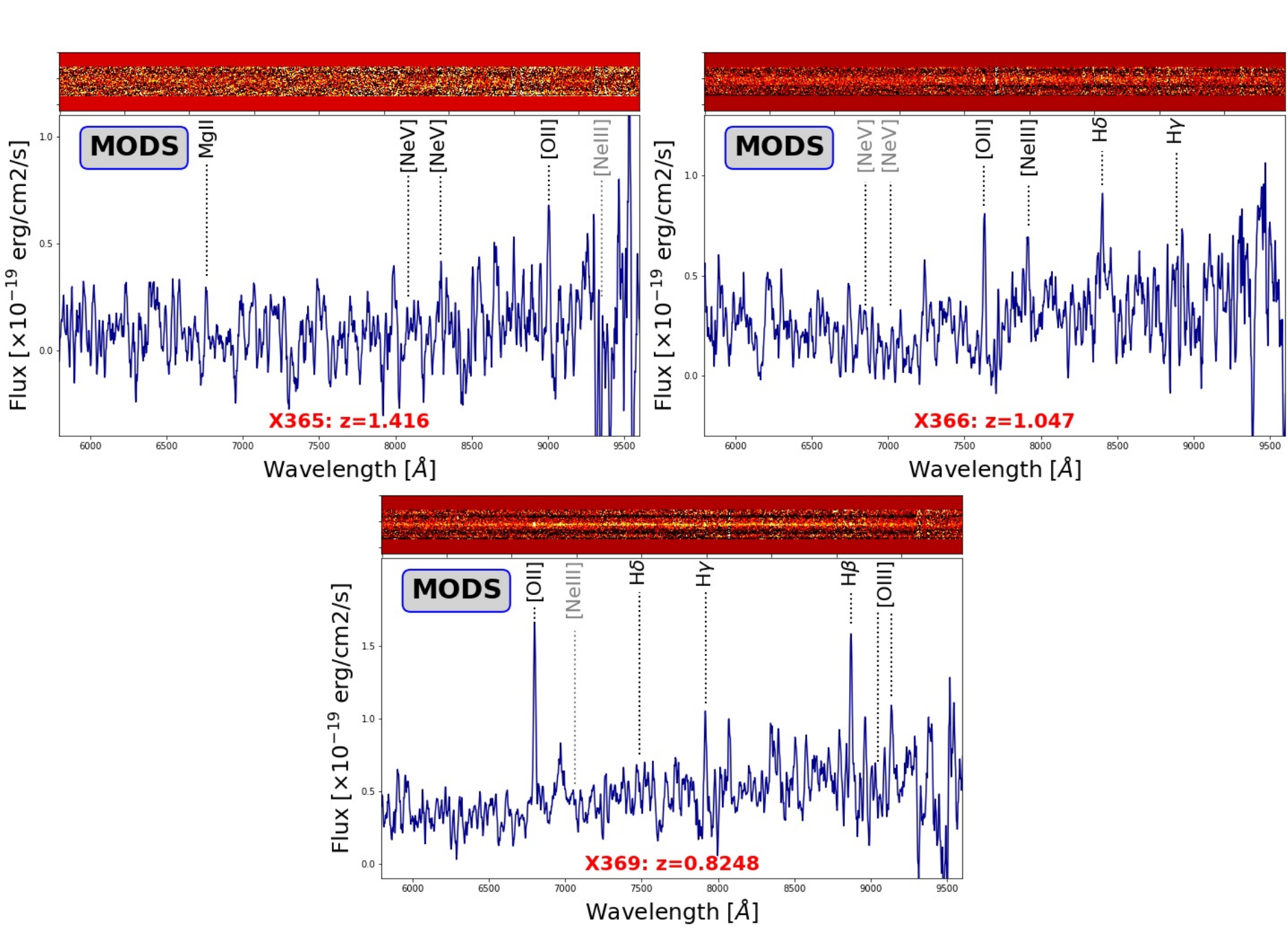}
\caption{\normalsize LBT-MODS
flux-calibrated spectra of three \cha\ J1030 sources targeted by the campaign presented in this work and not belonging to the $z\sim$2.78 structure. We highlighted the expected positions of the main emission lines that fall in the observed spectral range. On top of each one-dimensional spectrum, we report the two-dimensional one.
}\label{fig:spec_MODS_b}
\end{figure*}

We report in this Appendix, in Figures~\ref{fig:spec_MODS_a} and \ref{fig:spec_MODS_b}, the flux-calibrated spectra for the nine sources targeted by our LBT-MODS campaign for which we computed a spectroscopic redshift and that are not part of the $z\sim$2.78 LSS; the spectra of the LSS are reported in Figure~\ref{fig:spec_LSS}. The FITS file of these spectra, as well as of those of the six \cha\ J1030 for which we could not measure a spectroscopic redshift, are available online\footnote{\url{http://j1030-field.oas.inaf.it/~LBTz6/1030/xray_redshift_J1030.html}}.

\end{document}